\documentclass[%
 reprint,
nofootinbib,
 amsmath,amssymb,
 aps,
 prd,
floatfix,
]{revtex4-2}
\usepackage{graphicx}
\usepackage{dcolumn}
\usepackage{bm}
\usepackage{color}
\usepackage{amsmath}

\usepackage{amsmath,amsthm,amssymb}
\def\ket[#1]{{\ensuremath{|#1\rangle}}}
\usepackage[english]{babel}
\usepackage{url}
\usepackage{lineno}

\begin{document}

\preprint{APS/123-QED}

\title{New constraints on CPT symmetry violation in charm mesons
}

\author{Wojciech Krzemie\'n}
\author{Mateusz Kmieć}%
\author{Adam Szabelski}%
\author{Wojciech Wi\'slicki}

\affiliation{ 
National Centre for Nuclear Research, Otwock-Świerk, Poland
}




\date{\today}

\begin{abstract}
Small CPT-violating effects are admissible in several theories beyond the Standard Model of interactions. We analyse experimental data obtained with neutral flavoured meson mixing and place bounds on the scale of such phenomena. 
The constraints on the CPT-violating parameter $z$ are derived by reinterpreting the LHCb measurement of the time-dependent asymmetry in the Cabbibo-favoured  $D^0 \rightarrow K^{-}\pi^{+}$ and $\overline{D^{0}} \rightarrow K^{+}\pi^{-}$ decay rates.
The bounds provided in this paper are two orders of magnitude stricter compared to the results obtained by the FOCUS Collaboration. Among other constraints, we report bounds on the differences in the decay widths and masses of the neutral D meson flavour states at a 95\% confidence level. The decay width difference is bounded by $(-4.7<\delta\Gamma_D<2.1)\times 10^{-16}$GeV. The mass difference lies within~$(-2.0<\delta m_D<2.0)\times 10^{-15}$GeV for 95\% of $\text{arg}(z)$ values. The paper includes an assessment of the influence of CP violation on the results and a discussion of the future prospects. 
\end{abstract}

\maketitle

\section{Introduction}
\label{sect:introduction}
The CPT symmetry is defined as invariance under the joint transformation of charge conjugation C, spatial inversion P, and time-reversal T. It is one of the few fundamental symmetries that seem to be exactly conserved by Nature.
The profound significance of CPT symmetry is manifested in the {\it CPT theorem} (for a complete account cf. Ref. \cite{Streater:1989vi}), which states that all quantum field theories (QFTs) formulated on the flat space-time and respecting locality, unitarity and Lorentz invariance of interactions must conserve CPT symmetry. 
Hence, all QFTs describing fundamental particle interactions within the Standard Model (SM) must be CPT-invariant.

The importance of CPT symmetry is due to its deep connections to the Poincar\'e invariance of QFT in flat space-time. However, when gravitational effects are taken into account, global
spacetime symmetries are not symmetries of QFT anymore and relations between the CPT- and Lorentz invariance might have to be revisited. 
Therefore some extensions of QFT, incorporating gravity or strings, will also include violation of CPT at the scales of energy or space-time dimensions where quantum gravity (QG) is expected to be significant, i.e. close to the Planck scale.
At energies presently available at high-energy accelerators, or even in cosmic rays, these effects are expected to be tiny, and their observation requires significant advancements in detector sensitivity and accuracy.

Possible extensions of the SM can be studied in a model-independent framework based on the effective field theory, known as the Standard-Model Extension (SME) (cf. Refs. \cite{ colladayTestingCPTInvariance1995, colladayCPTViolationStandard1997}, and \cite{lehnertCPTLorentzSymmetry2023} for review).
The Lagrangian density in SME contains the CPT- and Lorentz-violating terms with coupling constants serving as control parameters contributing to several physical observables which values can be determined experimentally.

Many theoretical models of QG  explicitly incorporate CPT violation (CPTV), either by abandoning some assumptions of the CPT theorem or by the inclusion of the specific model of gravity, as e.g. the space-time deformation in non-commutative QFT at the Planck scale, where physical interactions are dominated by QG \cite{Amelino-Camelia:2000stu}.  
This motivates the construction of models connecting the dynamical or Lorentz-induced CPTV to many fundamental differences between matter and antimatter \cite{mavromatosSpontaneousCPTViolation2018}. For example, they predict observable effects in the differences of the particle and antiparticle lifetimes \cite{Bevilacqua:2022fbz} or in interference patterns of neutral mesons \cite{Bernabeu:2003ym,Bevilacqua:2024jpy}.
 An idea underlying these approaches is that CPTV revises the very concept of antiparticle and introduces terms with a wrong CPT parity to the two-particle states.

Apart from the explicit CPTV in fundamental interactions, there is another general mechanism of accomplishing this effect related to the emergence of the thermodynamic arrow of time in systems propagating in a dissipative environment.
This phenomenon can be studied in the density matrix formalism by following the time evolution of the density matrix expressed by the Gorini–Kossakowski–Sudarshan–Lindblad (GKSL) equation \cite{Gorini:1975nb, Lindblad:1975ef}, which differs from the usual von Neumann formula by an additional, dissipative term. 
The GKSL evolution scheme results in a decoherence of two- or many-particle quantum states which should be observable with a sensitive detector.
Such an effect, however, does not mean by itself that elementary interactions leading to decoherence violate CPT.
A deeper interest in this phenomenon is again connected to QG, as being a universal, irreducible background of micro black holes at the Planck scale.
These ideas were proposed theoretically in Refs.~\cite{waldQuantumGravityTime1980,hawkingUnpredictabilityQuantumGravity1982} and further developed in Refs. \cite{Ellis:1983jz,Huet:1994kr} as a model suitable for a coherent pair of neutral, heavy-flavoured mesons. 

In fundamental interactions, these effects can be experimentally approached at least in two ways. 
We can either study CPTV in the classical framework or in the effective field theory SME approach ~\cite{ colladayTestingCPTInvariance1995, colladayCPTViolationStandard1997, kosteleckyDataTablesLorentz2011}. In the classical method, the CPTV parameter is assumed to be a complex constant, whereas in the SME, it exhibits a momentum- and sidereal time dependence. In the latter approach, we are searching for CPTV induced by Lorentz-invariance violation. 
An extensive overview of SME-based tests of CPT symmetry can be found in Ref. ~\cite{lehnertCPTLorentzSymmetry2023}. In this paper, we will stay within the phenomenological and classical framework where the exact source of CPTV is not directly specified.

The consequence of the CPT theorem is that the basic properties of a particle and its antiparticle such as masses, lifetimes, and magnetic momenta must be the same. 
This leads to experimental tests of CPT invariance in which one looks for tiny differences between matter and antimatter.
Over the years, the searches for CPTV have been carried out with different experimental setups~\cite{linkCharmSystemTests2003,alicecollaborationPrecisionMeasurementMass2015,bevanTestingDiscreteSymmetries2016,moskalTestingCPTSymmetry2021, babusciDirectTestsCP2023, caloniProbingLorentzviolatingElectrodynamics2023, mishraSearchLorentzviolationSidereal2023},
while new approaches have been developed~\cite{karanUsingTimedependentIndirect2018,karanDealingCPTViolations2020, lehnertBetadecaySpectrumLorentz2022, vargasProspectsTestingLorentz2024, belyaev2024probingcptinvariancequarks, nowak2024cptlorentzsymmetrytests}.
So far, no experimental evidence of CPTV has been found.
Direct measurements are often experimentally difficult since strong and electromagnetic interactions dominate renormalised masses and lifetimes. 
Attractive possibilities are provided for CPT invariance tests in the neutral flavour mixing systems, such as neutrino and the neutral flavoured meson mixing dominated by the second-order weak interactions.   
Various CPTV scenarios and experimental tests with flavoured neutral mesons have been proposed ~\cite{Bernabeu:2003ym,bernabeuDecoherenceInducedViolation2006,edwardsSearchingCPTViolation2019,huangViolatingParametersDetermined2014,karanDealingCPTViolations2020, karanUsingTimedependentIndirect2018, 
shiExactResultsCP2013,shiExactTheoremsConcerning2012,
Bevilacqua:2022fbz, colladayTestingCPTInvariance1995,kosteleckySensitivityCPTTests1998, kosteleckySignalsCPTLorentz1999,robertsTestingSymmetryCorrelated2017}.

In this letter, we provide new bounds for CPTV in decays of charmed mesons. 
These constraints were derived by reinterpreting experimental measurements of CP violation (CPV) and mixing in the neutral D meson system. More specifically, we reanalysed the published LHCb results of the time-dependent asymmetry measurement in the Cabbibo-favoured (CF) $D^0 \rightarrow K^{-}\pi^{+}$
\footnote{Throughout this paper we assume the reaction is accompanied by its conjugate unless stated otherwise.}
decays \cite{aaijSearchTimedependentViolation2021}.
The constraints reported in this paper supersede the current bounds obtained by the FOCUS Collaboration~\cite{linkCharmSystemTests2003,particledatagroupReviewParticlePhysics2022}. 

The rest of this paper is structured as follows.
Section~\ref{sect:formalism} describes the mixing formalism for the neutral flavoured mesons including the parametrisation of the potential CPTV. Furthermore, the equations for the time-dependent asymmetry for uncorrelated neutral-flavoured mesons are provided. 
Section~\ref{sect:charm}  includes a review of the existing experimental limits on CPTV. 
In Section~\ref{sect:bounds}, the new bounds are presented, and the influence of the potential SM-compliant CPV effect on the obtained bounds is considered. Section~\ref{sect:discussion} is dedicated to future prospects in the context of the ongoing data-taking campaign at the Large Hadron Collider. Finally, alternative analysis approaches such as SME framework and interferometric measurements are briefly discussed. 

\section{Neutral flavoured meson time evolution formalism}
\label{sect:formalism}
In this Section, the formalism describing the time evolution of the neutral flavoured mesons is introduced. It includes mixing effects and the parametrisation of the CP, T and CPT violation. 

\subsection{The neutral meson system}    

The neutral meson system is described by a linear combination of $\ket[P^{0}]$ and $\ket[\overline{P^{0}}]$ strong interaction eigenstates (flavour states). The state of the system can be represented by the ket $\ket[\Psi(t)]$. In the Weisskopf–Wigner approximation (WWA) \cite{weisskopfBerechnungNaturlichenLinienbreite1930} the evolution of this state is governed by an effective hamiltonian $H^{\text{eff}}$, represented by a $2\times 2$ matrix, according to:

\begin{equation}
    i \hbar\frac{d}{dt}\ket[\Psi(t)]=H^{\text{eff}}\ket[\Psi(t)]=\left(M-\frac{i\Gamma}{2}\right)\ket[\Psi(t)],
    \label{shroedinger_Heff}
\end{equation}

\noindent where $\Gamma$ and $M$ are hermitian matrices. Here, $\Gamma$ represents the exponential decay component and $M$ is the mass term.
In this formalism, weak interactions are considered as a perturbation to the sum of the strong and electromagnetic parts of the Hamiltonian. The time evolution of flavour states is evaluated up to the second order in weak interactions and $H^{\text{eff}}$ is constructed to ensure the same evolution in the leading order of perturbation theory \cite{sozzimarcos.DiscreteSymmetriesCP2008}. 
The physical propagating states are the eigenstates of the hamiltonian $H^{\text{eff}}$ and denoted as $\ket[P_{1,2}(t)]$, henceforth referred to as the mass states. 
They evolve in time as:

\begin{equation}
   \ket[P_{1,2}(t)]=e^{-i\lambda_{1,2}t}\ket[P_{1,2}],
    \label{eq:eigenfunction2}
\end{equation}

\noindent where the complex parameters $\lambda_{1,2}$ represent the eigenvalues of $H^{\text{eff}}$, while $\ket[P_{1,2}]\equiv\ket[P_{1,2}(t=0)]$.
Since $\ket[P_{1,2}]$ have well defined masses $m_{1,2}$ and decay widths $\Gamma_{1,2}$,  $\lambda_{1,2}$ can be decomposed and represented as:
\begin{equation}
\lambda_{1,2}=m_{1,2}-i\Gamma_{1,2}/2.
\end{equation}

\subsection{Parameterisations of CPT and CP violation}

The mass states can be expressed as linear combinations of flavour states.
In this representation, effects of the CP, T and CPT violation can be naturally parametrised by introducing different, complex weights for the time-independent flavour eigenstates $\ket[P^{0}]$ and $\ket[\overline{P^0}]$ representing the particle and antiparticle \cite{sozzimarcos.DiscreteSymmetriesCP2008}:

 \begin{align} 
 \ket[P_{1}] &= N_{1}e^{i\eta_{1}}\left\{\ket[P^{0}] + \sqrt{\frac{1+z}{1-z}}\frac{q}{p}\ket[\overline{P^{0}}]\right\}, \nonumber \\
\ket[P_{2}] &= N_{2}e^{i\eta_{2}}\left\{\ket[P^{0}] - \sqrt{\frac{1-z}{1+z}}\frac{q}{p}\ket[\overline{P^{0}}]\right\}, 
\label{eq:mass_mixing}
\end{align}

\noindent where, $N_{1,2}=\left\{1+\left| \sqrt{\frac{1\pm z}{1\mp z}}\frac{q}{p} \right|^{2}\right\}^{-1/2}$ are the normalisation terms while $\eta_{1,2}$ are  phases that can be chosen freely.  
In this formalism, $p$ and $q$ parameters control T violation, with $\left|\frac{q}{p}\right|=1$ if and only if T is preserved. The parameter $z$ controls CPTV, with $z=0$ if and only if CPT is conserved. 
Both conditions need to be fulfilled for the CP symmetry to hold.

The time evolution of the flavour eigenstates can be explicitly accounted for by using Eqs. (\ref{eq:eigenfunction2},\ref{eq:mass_mixing})
\begin{align}
    \ket[P^{0}(t)]&= \left(g_{+}(t)+zg_{-}(t)\right)\ket[P^{0}]-\sqrt{1-z^{2}}\frac{q}{p}g_{-}(t)\ket[\overline{P^{0}}],\nonumber\\
    \ket[\overline{P^{0}}(t)]&= \left(g_{+}(t)-zg_{-}(t)\right)\ket[\overline{P^{0}}]-\sqrt{1-z^{2}}\frac{p}{q}g_{-}(t)\ket[P^{0}],
    \label{eq:time_evol} 
\end{align}

\noindent where $g_{\pm}(t) = \frac{1}{2}\left(e^{-im_{2}t-\Gamma_{2}t/2} \pm e^{-im_{1}t-\Gamma_{1}t/2}\right)$. 
The time-dependent decay probabilities of the flavour eigenstates (\ref{eq:time_evol}) to the final states $\ket[f]$ or the C-coupled $\ket[\overline f]$ can now be calculated as $P_{f}(t) = |\langle f | T | P^{0} \rangle|^{2}$, $ \overline{P}_{\overline{f}}(t) = |\langle \overline{f} | T | \overline{P^{0}} \rangle|^{2}$.

Matrix elements of the effective hamiltonian can be expressed using 
the $p,q,z$ parameters \cite{kosteleckySensitivityCPTTests1998}:

\begin{equation} \label{eq:H_eff_matrix}
H^{\text{eff}} = \frac{1}{2}\Delta\lambda\left(
\begin{aligned}
    \lambda/\Delta\lambda+z &~& \sqrt{1-z^2}q/p\\
    \sqrt{1-z^2}p/q &~& \lambda/\Delta\lambda-z
\end{aligned}
\right),
\end{equation}

\noindent where $\lambda = \lambda_{1}+\lambda_{2}$ and $\Delta\lambda = \lambda_{2}-\lambda_{1}$.

As discussed in Refs.~\cite{alankosteleckyExpectationValuesLorentz1996, sozzimarcos.DiscreteSymmetriesCP2008}, CPTV occurs if and only if the diagonal elements of $H^{\text{eff}}$ differ, whereas T violation requires that the moduli of the off-diagonal elements are different.
Using Eq. (\ref{eq:H_eff_matrix}), one can parametrise these two conditions as: 
\begin{align}
H^{\text{eff}}_{11}-H^{\text{eff}}_{22} &= z \Delta \lambda \nonumber \\
\left|H^{\text{eff}}_{12}\right|-\left|H^{\text{eff}}_{21}\right| &= \left|\frac{\Delta \lambda}{2}\sqrt{1-z^2}\right|(\left|q/p\right|-\left|p/q\right|)
\label{eq:Heffdiff}
\end{align}
and express the two parameters quantifying CPTV and T violation, $z$ and $q/p$, via elements of the effective hamiltonian $H^{\text{eff}}$ and its eigenvalues\footnote{$\delta m= m_{P^0}-m_{\overline{P^0}}$ and $\delta\Gamma = \Gamma_{P^0}-\Gamma_{\overline{P}^0}$ represent the differences of the diagonal elements of $M$ and $\Gamma$ matrices.}:
\begin{align}
    z &= \frac{\delta m - i \delta \Gamma/ 2 }{\Delta \lambda}, \nonumber \\
    q/p &= \frac{i\Im(M_{12})-\frac{1}{2}\Im(\Gamma_{12})}{\Delta\lambda},
    \label{zpq}
\end{align}

\noindent where $\Delta m$ and $\Delta \Gamma$ are mixing parameters defined as differences in mass and decay rates between the two mass states. 
They can be written in the dimensionless form, suitable for the analysis of $D^0$ mesons, as:

 \begin{equation} \label{eq:xy_def}
    x= \frac{\Delta m}{\Gamma},
    ~~~~~~~~~y=\frac{\Delta \Gamma}{2 \Gamma},
 \end{equation}
 where $\Delta m = m_2 - m_1$, $\Delta \Gamma = \Gamma_2 - \Gamma_1$ and $\Gamma =\frac{\Gamma_1 + \Gamma_2} {2}$ 
 is the average decay width.
The introduced formalism applies to all neutral flavoured pseudoscalar mesons. 
In particular, it can be used to describe the time evolution of the $D^{0}$ meson system.

\subsection{Time-dependent asymmetries}
It can be shown that for studies of  CPTV with the uncorrelated flavoured neutral mesons, the relevant observable is the time-dependent asymmetry, which can be constructed by combining the time-dependent decay probabilities in the following way~\cite{kosteleckyCPTLorentzViolation2001,vantilburgStatusProspectsCPT2015}: 
\begin{equation}
 A_{CPT}(t) = \frac{\overline{P}_{\overline{f}}(t) - P_f(t)}{\overline{P}_{\overline{f}}(t) + P_f(t)}.
\label{eq:ACPT_general}
\end{equation}
If we restrict our attention to flavour-specific final
states ($P_{\overline f}= 0 $, $\overline P_f = 0$) 
Eq.~\ref{eq:ACPT_fs}  simplifies to:
\begin{align}
   &A_{\text{CPT}}(t) =\nonumber\\& ~~A_{\text{dir}} + \frac{2\Re(z)\sinh\Delta\Gamma t/2 - 2\Im(z)\sin\Delta mt}{(1+|z|^2)\text{cosh}\Delta\Gamma t/2 + (1-|z|^2)\text{cos}\Delta m t},
\label{eq:ACPT_fs}
\end{align}
where all asymmetries, including the direct CPTV term $A_{\text{dir}}$,
 are assumed to be much smaller than one. In this case, the measurement of the CPTV is independent of the CPV effects. 

The oscillation in the neutral charm meson system is very slow compared to the $D^0$ decay time $\tau$, since the corresponding mixing parameters $x$ and $y$ are of the order of $\sim 0.5 \%$~\cite{amhisAveragesHadronHadron2023}. This allows us to further simplify \eqref{eq:ACPT_fs} by approximating it to the leading order in these parameters ($x \tau \Gamma \ll 1$ and $y \tau \Gamma \ll1$): 

\begin{equation}
\label{eq:ACPTlinear}
  A_{\text{CPT}}(t) = A_{\text{dir}} + (\Re(z) y - \Im(z) x ) \Gamma t   , 
\end{equation}

The CPTV term in Eq. \eqref{eq:ACPTlinear} corresponds to the slope of the line describing the dependence of the asymmetry on decay time.

\subsection{Influence of CP violation on CPT asymmetry}
\label{sect:CP_influence} 
Studies presented here are based on the two-hadron decays of $D^0$ mesons: $D^0\rightarrow K^-\pi^+$ and its CPT-conjugate $\overline{D^0}\rightarrow K^+\pi^-$.
Decaying mesons originate from the strongly-decaying $D^\ast$ mesons: $D^{{*}{+}}(2010)\rightarrow D^0\pi^{+}$ and $D^{{*}{-}}(2010)\rightarrow \overline{D^0}\pi^{-}$. 
Electric charge of the low-momentum (so-called {\it soft}) pion indicates the initial flavour, while the final flavour is determined by the pion charge from the $D^0$ decay.
The soft and final pions are used for flavour tagging.
When both tagging pions are of the same sign the flavours of the initial and final state coincide. 
An asymmetry determined from such events is called the right-sign (RS) asymmetry and is equal to
\begin{equation}
  A_{RS}(t) = \frac{N(D^0\to K^-\pi^+)(t)-N(\overline{D^0}\to K^+\pi^-)(t)}{N(D^0\to K^-\pi^+)(t)+N(\overline{D^0}\to K^+\pi^-)(t)}\label{eq:A_RS}.
\end{equation}
The $A_{RS}$ asymmetry (\ref{eq:A_RS}) is dominated by the direct CF decays of $D^{0}$ mesons, which makes it almost fully flavour-specific. 
The only contamination is due to the cases where $D^{0}$ mesons oscillate into $\overline{D^0}$ and subsequently decay  via a Doubly-Cabbibo suppressed (DCS) mode $\overline{D^0}\rightarrow K^{-} \pi^{+} $. 
However, the smallness of $D^0\rightarrow K^{+} \pi^{-} $ branching fraction, $1.36 \times 10^{-4}$ for DCS as compared to $3.947 \times 10^{-2}$ ~\cite{particledatagroupReviewParticlePhysics2022}  for CF, combined with the slow mixing (oscillation period is three orders of magnitude larger than $D^0$ decay time) makes this admixture so small that it is usually neglected in the analyses. 
Nevertheless, with the increasing experimental precision these phenomena could become important. 
Indeed, the potential effects of mixing and time-dependent CPV in $D^0\rightarrow K^{+}\pi^{-}$ channel are discussed in detail in Refs.~\cite{pajeroMixingCPViolation2022, schwartzEffectBarMixing2022} in the context of $y_{CP}$ measurements.   

In our case, the possible SM-compliant CPV would manifest as an additional linear time-dependent term in the asymmetry that could mimic the CPTV effect. 
Taking into account the DSC contribution the asymmetry in Eq. (\ref{eq:ACPTlinear}) becomes:
\begin{equation}
\begin{aligned}
&A_{\text{CPT}}(t) =  A_{\text{dir}} + (\Re(z) y - \Im(z) x )\Gamma t\\&-\frac{\sqrt{R_{DCS}}}{2}\sqrt{|1-z^2|}\\
&\times\Big\{\sin{\phi}\left[x\cos{(\delta+\kappa)}-y\sin{(\delta+\kappa)})\right]\left(\left|\frac{q}{p}\right|+\left|\frac{p}{q}\right|\right)\\
&-\cos{\phi}\left[y\cos{(\delta+\kappa)}+x\sin{(\delta+\kappa)})\right]\left(\left|\frac{q}{p}\right|-\left|\frac{p}{q}\right|\right)\Big\}\Gamma t,
\end{aligned}
\label{eq:DCS}
\end{equation}
 where $\kappa \equiv 0.5\,\text{arg}(1-z^2)$, $R_{DCS} \equiv |A_{DCS}/A_{CF}|^{2}$, weak phase difference $\phi \equiv  \text{arg}(q/p)$, and the strong phase difference $\delta \equiv\text{arg}(A_{CF}/A_{DCS})$.
Since both CF and DCS decays in the SM proceed via tree-level amplitudes dominated by a single weak phase, the direct CPV term can be
neglected~\cite{particledatagroupReviewParticlePhysics2022}. Generally, the mixing and CPV effects in charm are strongly suppressed in the SM predictions, however, they are rather imprecise because of the difficulty in calculation of contributions dominated by the long-distance interactions~\cite{particledatagroupReviewParticlePhysics2022}.

\section{Experimental study of CPT violation in charm decays}
\label{sect:charm}

\begin{figure}[tbh]
    \centering
    \includegraphics[width=0.48\textwidth]{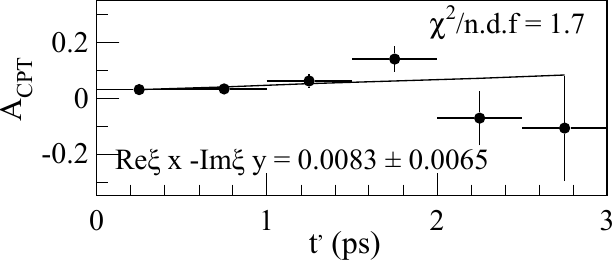}
    \includegraphics[width=0.48\textwidth]{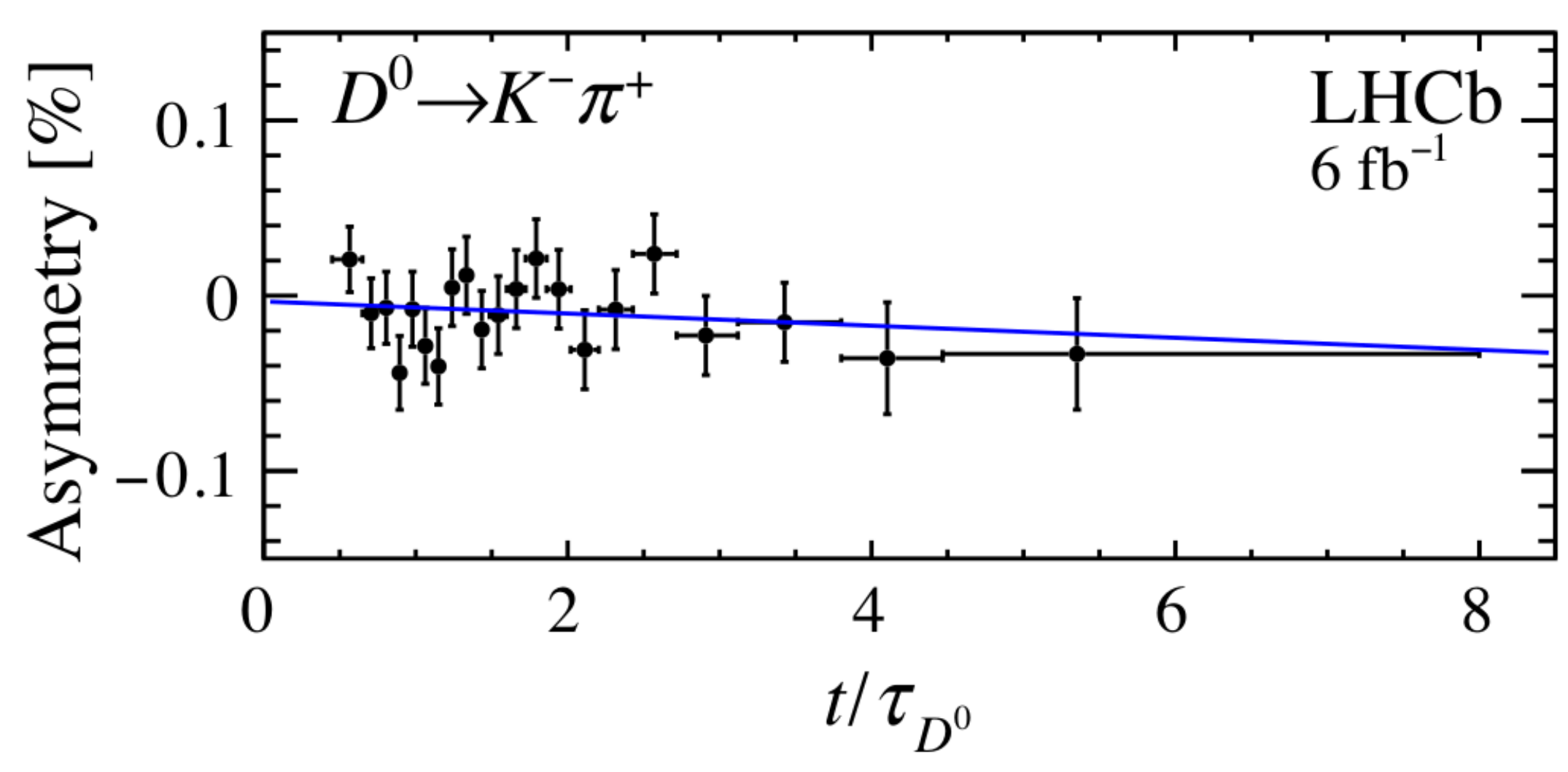}
    \caption{Experimental, time-dependent $A_{RS}$ asymmetry of $D^0 \rightarrow K^{-}\pi^{+}$ candidates determined by the FOCUS (top plot), and by the LHCb (bottom plot) Collaborations. The linear fit is superimposed in the bottom plot with the $\chi^{2}\slash\mbox{ndf}$ per degree of freedom equal to $17\slash 19$. The figures are adapted from Refs. ~\cite{linkCharmSystemTests2003} and~\cite{aaijSearchTimedependentViolation2021}, respectively.} \label{fig:exp_asymmetry}
\end{figure}

The RS asymmetry $A_{RS}$ (\ref{eq:A_RS}) is an experimental observable used in our studies.
The first experimental search for the CPTV was performed by the FOCUS Collaboration~\cite{linkCharmSystemTests2003}.
The charm mesons were produced in the interaction of the 180 GeV photons with the fixed target.
The $A_{RS}(t)$ is presented in Fig.~\ref{fig:exp_asymmetry} (upper plot). 
Due to the limited event sample consisting of  
35 thousand event candidates, the estimated 
limits for the expression $\Re(z) - \Im(z)$ provided rather loose bounds of the order $\mathcal O(1)$ but till now representing the best CPT upper limit in the charm sector.
The FOCUS analysis was also performed in the SME (\cite{kosteleckySensitivityCPTTests1998}) framework.
Due to its abundance the CF mode $D^{0}\rightarrow  K^{-} \pi^{+}$ was used as control channel by several analyses performed by the LHCb Collaboration~\cite{aaijObservationMassDifference2019, aaijUpdatedmeasurment2020, aaijTimeIntegrated2023}.
In particular, the RS asymmetry was determined as a reference observable (see Fig.~\ref{fig:exp_asymmetry} bottom plot) in the analysis of the asymmetries in the time-dependent rates of $D^0\to K^+K^-$ and $D^0\to\pi^+\pi^-$ decays
measured in a $pp$ collision data sample collected with the LHCb detector~\cite{aaijSearchTimedependentViolation2021}. 

In this paper, we use the $A_{RS}$ to extract bounds on the CPTV parameter $z$. Due to the abundance of the LHCb sample and the good control of systematic errors, we can significantly improve the accuracy of the CPT invariance.

\section{New bounds on CPT violation in the charm sector}
\label{sect:bounds}

According to the model defined in Eq.~(\ref{eq:ACPTlinear}), the slope of the linear fit to the time-dependent RS asymmetry (Fig.~\ref{fig:exp_asymmetry} bottom plot) can be identified as the $y\Re(z) -x\Im(z)$ term. 
The value of the slope determined from the fit to data is equal to
$s=(-4\pm5\pm2) \times 10^{-5}$, where the uncertainties correspond to statistical and systematic, respectively~\cite{aaijSearchTimedependentViolation2021}.  
The result is consistent with zero within one statistical standard deviation.  
The real and imaginary parts of $z$ cannot be disentangled from this fit without further assumptions. 

\begin{figure}[tbh]
    \centering
    \includegraphics[width=0.9\linewidth]{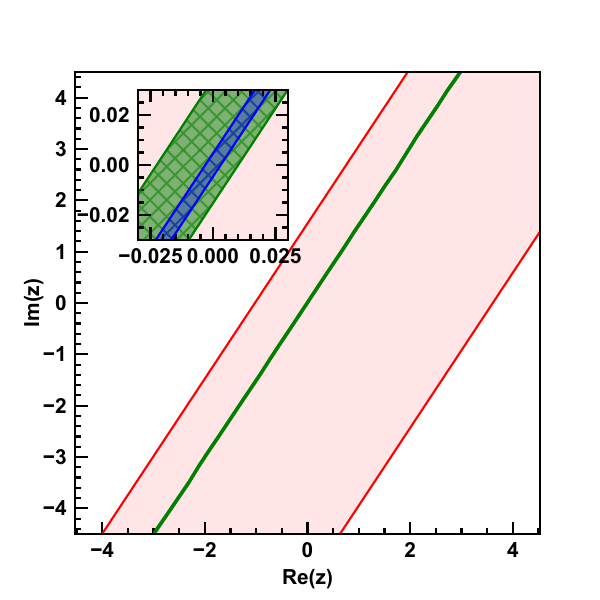}
    \caption{Experimental limits on the CPT-violating $z$ parameter at the 95 \% CL in the $D^0 \rightarrow K^{-}\pi^{+}$ channel by the FOCUS Collaboration~\cite{linkCharmSystemTests2003} and deduced from the LHCb measurement \cite{aaijSearchTimedependentViolation2021} denoted as red and green area respectively. The uncertainty originating from the possible SM-compliant CPV is shown as a blue area.}
    \label{fig:limits}
\end{figure}

Using the value of $s$, one gets a linear dependence of the imaginary and real parts of $z$ as $\Im{(z)} = y\Re{(z)}/x - s/x$ in the $(\Re(z), \Im(z))$ plane. 
The determined constraint region at 95 \% confidence level (CL), is presented in Fig.~\ref{fig:limits} as a green area, where the statistical and systematic uncertainties of all quantities were added in quadrature. 
The previous strongest bound established by FOCUS Collaboration is denoted as the red area and the possible SM-compliant CPV is shown in blue. 
The hypothetical influence of the CP violation is discussed in the next section.
The values of $x$ and $y$ mixing parameters are taken from HFLAV group~\cite{amhisAveragesHadronHadron2023}. 
The uncertainties of the mixing parameters are taken into account by assuming the Gaussian prior with the standard deviation set to the parameter uncertainties.  
We note that the production and detection asymmetries enter as a time-independent additive factor that does not affect the final slope~\cite{aaijSearchTimedependentViolation2021}.  

The parameters needed to estimate the effect of CPV are taken from the HFLAV fits~\cite{amhisAveragesHadronHadron2023} and gathered in Tab.~\ref{tab:params}. 
\renewcommand{\arraystretch}{1.5}
\begin{table}[tbh]
    \scriptsize
    \centering
    \begin{tabular}{|c|c|c|c|c|c|}\hline
        $x[\%]$ & $y[\%]$ & $\delta[^{\circ}]$ & $R_{DCS}[\%]$ & $|q/p|$ & $\phi[^{\circ}]$\\\hline
         \vspace{1pt}$0.407^{+0.044}_{-0.044}$& $0.645^{+0.024}_{-0.023}$ & $11.4^{+3.5}_{-3.8}$ & $0.344^{+0.002}_{-0.002}$ & $0.994^{+0.016}_{-0.015}$ & $-2.6^{+1.1}_{-1.2}$\\\hline
    \end{tabular}
    \caption{Parameters entering the CPV term in Eq. (\ref{eq:DCS}) taken from HFLAV fit \cite{amhisAveragesHadronHadron2023}.} 
    \label{tab:params}
\end{table}
Inspecting the values of the $\phi$ and $|q/p|$ one sees again that the expected CP effect is tiny but the uncertainties are rather large. 
Assuming that the CPTV is small, i.e. $|z|^{2}\ll 1$, and taking into account the HFLAV parameters with their uncertainties, we estimated the 95\%~CL interval for the CP-related term proportional to $\sqrt{R_{DCS}}$ in Eq.~\eqref{eq:DCS} to be $(-1.0 \pm 2.0)\times 10^{-5}$.

This result is consistent with the one determined in Ref.~\cite{aaijSearchTimedependentViolation2021}. 
The CP-related term in the $(\Re(z), \Im(z))$ space is represented as a blue band in Fig.~\ref{fig:limits}. 
At the current experimental accuracy level, the CPV term is still small, however, it can become significant in future analyses with higher statistics. 
On the other hand, with more collected data from LHCb, Belle-2 and BES-III, CPV bounds may be tightened by the independent measurements.

\section{Discussion and prospects}
\label{sect:discussion}
\begin{figure}[tbh]
    \centering
    \includegraphics[width=0.8\linewidth]{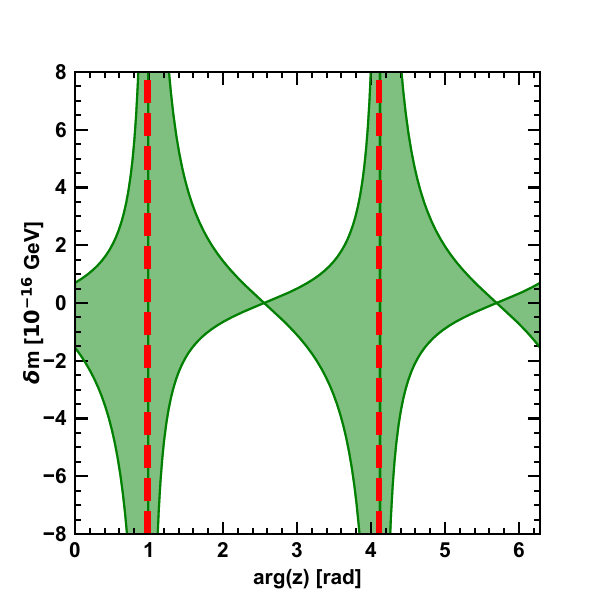}
    \caption{Distribution of the mass difference $\delta m$ parameter as a function of $z$ phase \eqref{dmargz}. The green area denotes the 95\%~CL. Red dashed lines show the $arg(z)$ values for which $\delta m$ is undefined since the linear asymmetry term vanishes.}
    \label{fig:deltam}
\end{figure}

The constraints on CPTV restrict the space of the possible values of particle-antiparticle mass and decay width differences, $\delta m$ and $\delta \Gamma$ respectively via Eq.~(\ref{zpq}). 
The slope of the RS asymmetry is directly connected to $\delta \Gamma$ (see appendix~\ref{appendix:relation}). This relationship can be used to constrain the decay width difference at 95\%~CL:  $(-4.7 <\delta \Gamma < 2.1)\times 10^{-16}$ GeV. In contrast, without further assumptions, $\delta m$ bounds cannot be deducted from the linear term of asymmetry. For instance, $\delta m$ can be expressed as a function of phase $\theta \equiv arg(z)$ (cf. Eq. \eqref{deltamphi}: 
\begin{equation} \label{dmargz}
\delta m(\theta) =s \times \frac{x \cos(\theta) + y  \sin(\theta)}{y \cos(\theta) -x \sin(\theta)}. 
\end{equation}
This dependence leads to $\delta m$ confidence intervals corresponding to the  95\%~CL shown as a green area in Fig.~\ref{fig:deltam}.
For the two angles where the linear term of asymmetry vanishes, $\delta m(\theta)$  is undefined. The singular points in Fig \ref{fig:deltam} are marked with red dashed lines. Outside this neighbourhood, the estimated $\delta m$ is bounded by values of the order of $10^{-15} - 10^{-16}$ GeV. More specifically, this mass difference lies within $(-2.0<\delta m<2.0)\times 10^{-15}$ GeV excluding 5\% of arg(z) corresponding to the largest $|\delta m|$ values. In particular, the assumption that $\Re(z)=0$ results in the following constraint: $ (-1.6<\delta m < 3.5)\times 10^{-16}$ GeV. 

\begin{figure}[tbh]
    \centering
    \includegraphics[width=\linewidth]{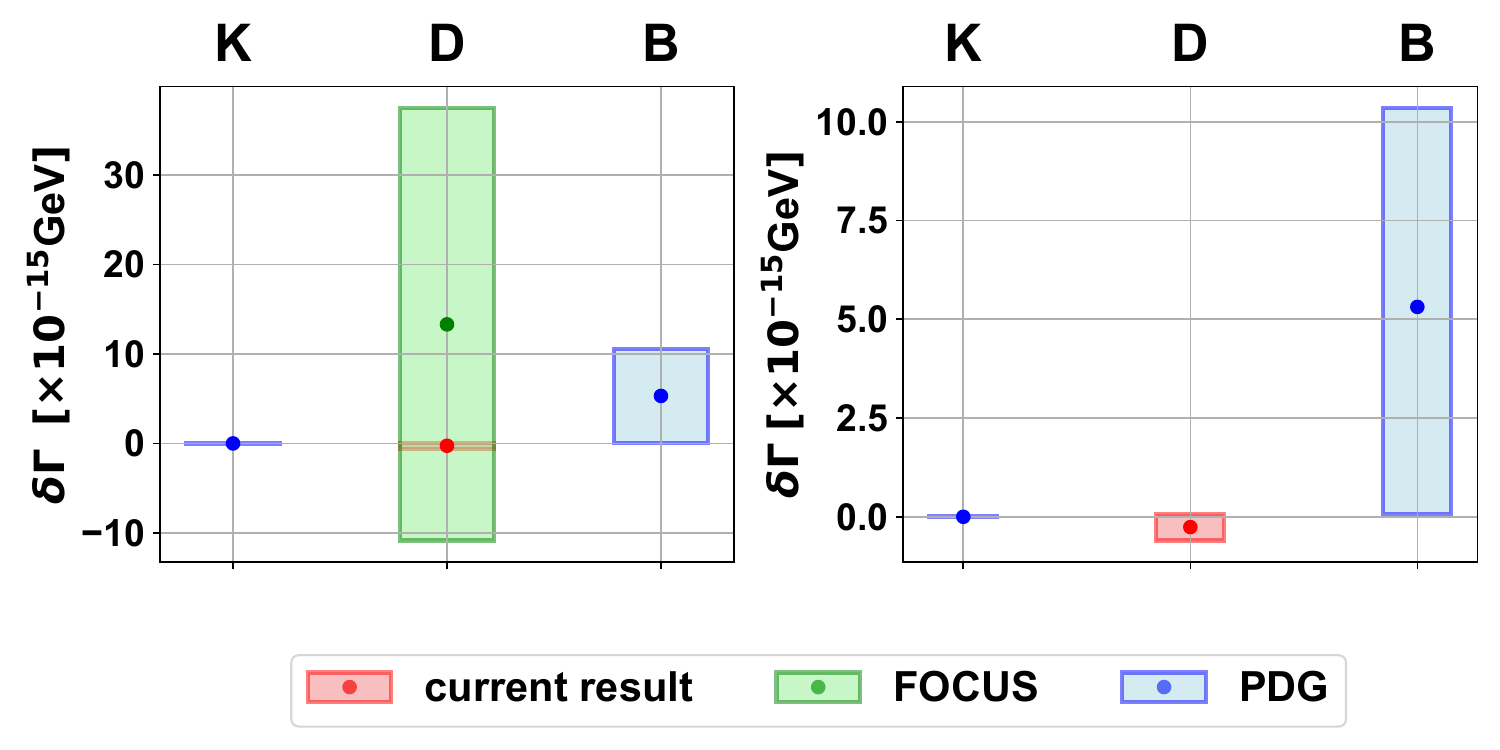}
    \caption{Current experimental bounds at \% 95 CL on particle-antiparticle decay width difference $\delta\Gamma$. The right-hand plot is the same as the left-hand one but on a smaller scale,  with the FOCUS result removed. In the case of $D$ and $K$ the direct measurement results are available while for $B$ the limits are obtained from $\Re(z)$ and $\Im(z)$ PDG averages according to the formula $\delta\Gamma = 2\Gamma[\Re(z)y-\Im(z)x]$. The bounds for $B_s$ are much broader and are not shown.} 
    \label{fig:deltaGamma}
\end{figure}

The new constraints in the charm sector can be juxtaposed with the current experimental limits on CPTV in other neutral flavoured meson systems~\cite{particledatagroupReviewParticlePhysics2022}. 
The experimental precision to $\delta m$ and $\delta \Gamma$ is determined by the interplay of the relative values of the decay time scale and the mixing parameters, which differ considerably between meson families as discussed in Section~\ref{sect:charm}. The strange sector is the most favourable from the point of view of CPTV studies. 
Indeed, the smallness of the denominator value in Eq.~(\ref{zpq}) for the neutral kaons, strongly amplifies sensitivity to $\delta m$ and $\delta \Gamma$ and results in a precision of about $10^{-18}$~GeV~\cite{angelopoulos1999}.
The comparison between 95\% CL limits for the decay width differences is shown in  Fig.~\ref{fig:deltaGamma}.  
The constraint on $\delta \Gamma$ in the $D^{0}$ system, deduced in this article, reaches $10^{-16}$ GeV. 
In the case of the bottom and bottom-strange systems,  no direct $\delta\Gamma$ measurements are available and their values are obtained from the combination of $\Re(z)$ and $\Im(z)$ measured by Belle, BaBar and LHCb Collaborations.
The result for $B$ is of the order of $\sim10^{-14}$~GeV \cite{Higuchi:2012kx, BaBar:2016zvy}, while for $B_s$ it reaches the value at the level of $\sim10^{-12}$~GeV \cite{aaijSearchViolationsLorentz2016}. 
The value of CPTV coupling can, in principle,  vary depending on the meson quark content, and therefore can be different for each flavour family~\cite{kosteleckyCPTLorentzViolation2001}.

In the ongoing data-taking period at the Large Hadron Collider, the collected statistics of 
$D^0 \rightarrow K^{-}\pi^{+}$ candidates  
should reach $3\times10^9$ \cite{LHCb:2018roe,run3fig}. 
Naive extrapolation of the statistical uncertainty for the charm CPT bounds gives a scale-down factor of $2.45$. 
At that moment, the statistical uncertainty would reach the systematic error level. 
The bulk part of the latter is statistics-dependent (cf. Ref. \cite{aaijSearchTimedependentViolation2021}) and, hopefully, could be further reduced. 
The uncertainty of the $DCS$ component, connected to CPV, should also be further constrained. 
However, 
as discussed in Section~\ref{sect:CP_influence}, the precision of the future CPT measurement in the charm sector is affected by the possible SM-compliant CPV effects, which would mimic the CPT behaviour and must be taken into account in future analyses. 
This can be done within the SME approach, by exploiting the SME degrees of freedom such as dependence on the particle boost or modulations in sidereal time, which are expected to be independent of CPV effects, this issue can be overcome~\cite{kosteleckySensitivityCPTTests1998}.  
A detailed discussion concerning the prospects of tests of CPT invariance within the SME framework lies outside of the scope of this article and will be presented in a separate review.  

The interferometric studies of the coherent $D^{0}$ pairs time evolution are an attractive alternative to the CPT tests with the uncorrelated mesons~\cite{colladayTestingCPTInvariance1995,kosteleckySignalsCPTLorentz1999,Shi:2016bvo}. 
Application of the filtering method based on the final-state flavour-  and CP tagging~\cite{Banuls:1999aj,Bernabeu:2012ab,Bernabeu:2012nu,Bernabeu:2015aga} would provide a way to separately test CP, T and CPT symmetries in a manner analogous to the measurements performed with B and K mesons~\cite{BaBar:2012bwc, BaBar:2016zvy,babusciDirectTestsCP2023}.
However, in experiments such as BES-III~\cite{BESIII:2009fln} where the quantum coherent $D^{0}$ pairs are copiously produced in the $\psi(3770)$ decay, lifetime analysis measurements are very challenging, due to low momenta in the laboratory frame of the  $D^{0}$ mesons produced in the symmetric $e^{+}e^{-}$ collisions. Hence, $D^{0}$ time-of-flight distances are impossible to measure.   
Such types of investigations will be feasible in future facilities operating at the asymmetric 
$e^{+}e^{-}$ colliders such as Super $\tau$-Charm facility~\cite{Achasov:2023gey}. 
Another interesting idea of quantum-correlated measurement is discussed in Ref.~\cite{naikNovelCorrelatedOverline2023} where it is proposed to exploit the quantum-correlated $D^{0}$ pairs
produced in  
$ \chi_{c1}(3872) \rightarrow D^0\overline{D^{0}}\pi^0$ decay. 
In this case,  the meson pair can be described by a symmetric, i.e. $C=1$ wave function, in contrast to the asymmetric $C=-1$ function describing the decay products of $\psi(3770)$. Analogously to the filtering method, a set of transition conjugate decay modes is proposed. The author suggests that the analysis would be feasible in the facilities operating today, such as LHCb or Belle-II. 
\section{Summary}
\label{sect:summary}
We have established new constraints on CPT violation in neutral charm meson mixing through a reinterpretation of the RS time-dependent experimental asymmetry in decay rates for $D^0 \rightarrow K^{-}\pi^{+}$ and $\overline{D^0} \rightarrow K^{+}\pi^{-}$.  
The determined 95\%~CL limit $(-14.6<\Re{(z)}y-\Im{(z)}x<6.6)\times 10^{-5}$ is two orders of magnitude stricter compared to the previous bounds set by FOCUS.
This result corresponds to precision of the order of $10^{-15} - 10^{-16}$ GeV
in terms of particle-antiparticle mass and decay time differences.
By taking the mixing parameters $x$ and $y$ from HFLAV one finds a 95\%~CL limit for the decay width difference equal to  $(-4.7<\delta \Gamma < 2.1 )\times 10^{-16}$ GeV. 
The bounds on the mass difference cannot be extracted without additional assumptions. For example, assuming $\Re(z)=0$, one finds $(-1.6 <\delta m < 3.5)\times 10^{-16}$ GeV at 95\%~CL. 
These results represent the most stringent constraint on CPTV in the charm sector.
The potential presence of SM-compliant CPV terms and their uncertainty can limit the sensitivity of future phenomenological CPT analyses using larger statistical samples. 
The analysis in the SME framework should be less susceptible to these effects.

\newcommand\noopsort[1]{} 
\bibliographystyle{apsrev4-1}
\bibliography{cpt_classical}

\begin{thebibliography}{65}%
\makeatletter
\providecommand \@ifxundefined [1]{%
 \@ifx{#1\undefined}
}%
\providecommand \@ifnum [1]{%
 \ifnum #1\expandafter \@firstoftwo
 \else \expandafter \@secondoftwo
 \fi
}%
\providecommand \@ifx [1]{%
 \ifx #1\expandafter \@firstoftwo
 \else \expandafter \@secondoftwo
 \fi
}%
\providecommand \natexlab [1]{#1}%
\providecommand \enquote  [1]{``#1''}%
\providecommand \bibnamefont  [1]{#1}%
\providecommand \bibfnamefont [1]{#1}%
\providecommand \citenamefont [1]{#1}%
\providecommand \href@noop [0]{\@secondoftwo}%
\providecommand \href [0]{\begingroup \@sanitize@url \@href}%
\providecommand \@href[1]{\@@startlink{#1}\@@href}%
\providecommand \@@href[1]{\endgroup#1\@@endlink}%
\providecommand \@sanitize@url [0]{\catcode `\\12\catcode `\$12\catcode `\&12\catcode `\#12\catcode `\^12\catcode `\_12\catcode `\%12\relax}%
\providecommand \@@startlink[1]{}%
\providecommand \@@endlink[0]{}%
\providecommand \url  [0]{\begingroup\@sanitize@url \@url }%
\providecommand \@url [1]{\endgroup\@href {#1}{\urlprefix }}%
\providecommand \urlprefix  [0]{URL }%
\providecommand \Eprint [0]{\href }%
\providecommand \doibase [0]{http://dx.doi.org/}%
\providecommand \selectlanguage [0]{\@gobble}%
\providecommand \bibinfo  [0]{\@secondoftwo}%
\providecommand \bibfield  [0]{\@secondoftwo}%
\providecommand \translation [1]{[#1]}%
\providecommand \BibitemOpen [0]{}%
\providecommand \bibitemStop [0]{}%
\providecommand \bibitemNoStop [0]{.\EOS\space}%
\providecommand \EOS [0]{\spacefactor3000\relax}%
\providecommand \BibitemShut  [1]{\csname bibitem#1\endcsname}%
\let\auto@bib@innerbib\@empty
\bibitem [{\citenamefont {Streater}\ and\ \citenamefont {Wightman}(1989)}]{Streater:1989vi}%
  \BibitemOpen
  \bibfield  {author} {\bibinfo {author} {\bibfnamefont {R.~F.}\ \bibnamefont {Streater}}\ and\ \bibinfo {author} {\bibfnamefont {A.~S.}\ \bibnamefont {Wightman}},\ }\href@noop {} {\emph {\bibinfo {title} {{PCT, spin and statistics, and all that}}}}\ (\bibinfo  {publisher} {Princeton University Press},\ \bibinfo {year} {1989})\BibitemShut {NoStop}%
\bibitem [{\citenamefont {Colladay}\ and\ \citenamefont {Kosteleck{\'y}}(1995)}]{colladayTestingCPTInvariance1995}%
  \BibitemOpen
  \bibfield  {author} {\bibinfo {author} {\bibfnamefont {D.}~\bibnamefont {Colladay}}\ and\ \bibinfo {author} {\bibfnamefont {V.~A.}\ \bibnamefont {Kosteleck{\'y}}},\ }\href {\doibase 10.1103/PhysRevD.52.6224} {\bibfield  {journal} {\bibinfo  {journal} {Phys. Rev. D}\ }\textbf {\bibinfo {volume} {52}},\ \bibinfo {pages} {6224} (\bibinfo {year} {1995})}\BibitemShut {NoStop}%
\bibitem [{\citenamefont {Colladay}\ and\ \citenamefont {Kostelecky}(1997)}]{colladayCPTViolationStandard1997}%
  \BibitemOpen
  \bibfield  {author} {\bibinfo {author} {\bibfnamefont {D.}~\bibnamefont {Colladay}}\ and\ \bibinfo {author} {\bibfnamefont {A.}~\bibnamefont {Kostelecky}},\ }\href {\doibase 10.1103/PhysRevD.55.6760} {\bibfield  {journal} {\bibinfo  {journal} {Phys. Rev. D}\ }\textbf {\bibinfo {volume} {55}},\ \bibinfo {pages} {6760} (\bibinfo {year} {1997})},\ \Eprint {http://arxiv.org/abs/hep-ph/9703464} {arxiv:hep-ph/9703464} \BibitemShut {NoStop}%
\bibitem [{\citenamefont {Lehnert}(2023)}]{lehnertCPTLorentzSymmetry2023}%
  \BibitemOpen
  \bibfield  {author} {\bibinfo {author} {\bibfnamefont {R.}~\bibnamefont {Lehnert}},\ }in\ \href@noop {} {\emph {\bibinfo {booktitle} {CPT and Lorentz Symmetry}}}\ (\bibinfo  {publisher} {WORLD SCIENTIFIC},\ \bibinfo {address} {Indiana University Bloomington, USA},\ \bibinfo {year} {2023})\BibitemShut {NoStop}%
\bibitem [{\citenamefont {Amelino-Camelia}(2002)}]{Amelino-Camelia:2000stu}%
  \BibitemOpen
  \bibfield  {author} {\bibinfo {author} {\bibfnamefont {G.}~\bibnamefont {Amelino-Camelia}},\ }\href {\doibase 10.1142/S0218271802001330} {\bibfield  {journal} {\bibinfo  {journal} {Int. J. Mod. Phys. D}\ }\textbf {\bibinfo {volume} {11}},\ \bibinfo {pages} {35} (\bibinfo {year} {2002})},\ \Eprint {http://arxiv.org/abs/gr-qc/0012051} {arXiv:gr-qc/0012051} \BibitemShut {NoStop}%
\bibitem [{\citenamefont {Mavromatos}\ and\ \citenamefont {Sarkar}(2018)}]{mavromatosSpontaneousCPTViolation2018}%
  \BibitemOpen
  \bibfield  {author} {\bibinfo {author} {\bibfnamefont {N.}~\bibnamefont {Mavromatos}}\ and\ \bibinfo {author} {\bibfnamefont {S.}~\bibnamefont {Sarkar}},\ }\href {\doibase 10.3390/universe5010005} {\bibfield  {journal} {\bibinfo  {journal} {Universe}\ }\textbf {\bibinfo {volume} {5}},\ \bibinfo {pages} {5} (\bibinfo {year} {2018})}\BibitemShut {NoStop}%
\bibitem [{\citenamefont {Bevilacqua}\ \emph {et~al.}(2022)\citenamefont {Bevilacqua}, \citenamefont {Kowalski-Glikman},\ and\ \citenamefont {Wislicki}}]{Bevilacqua:2022fbz}%
  \BibitemOpen
  \bibfield  {author} {\bibinfo {author} {\bibfnamefont {A.}~\bibnamefont {Bevilacqua}}, \bibinfo {author} {\bibfnamefont {J.}~\bibnamefont {Kowalski-Glikman}}, \ and\ \bibinfo {author} {\bibfnamefont {W.}~\bibnamefont {Wislicki}},\ }\href {\doibase 10.1103/PhysRevD.105.105004} {\bibfield  {journal} {\bibinfo  {journal} {Phys. Rev. D}\ }\textbf {\bibinfo {volume} {105}},\ \bibinfo {pages} {105004} (\bibinfo {year} {2022})},\ \Eprint {http://arxiv.org/abs/2201.10191} {arXiv:2201.10191 [hep-th]} \BibitemShut {NoStop}%
\bibitem [{\citenamefont {Bernabeu}\ \emph {et~al.}(2004)\citenamefont {Bernabeu}, \citenamefont {Mavromatos},\ and\ \citenamefont {Papavassiliou}}]{Bernabeu:2003ym}%
  \BibitemOpen
  \bibfield  {author} {\bibinfo {author} {\bibfnamefont {J.}~\bibnamefont {Bernabeu}}, \bibinfo {author} {\bibfnamefont {N.~E.}\ \bibnamefont {Mavromatos}}, \ and\ \bibinfo {author} {\bibfnamefont {J.}~\bibnamefont {Papavassiliou}},\ }\href {\doibase 10.1103/PhysRevLett.92.131601} {\bibfield  {journal} {\bibinfo  {journal} {Phys. Rev. Lett.}\ }\textbf {\bibinfo {volume} {92}},\ \bibinfo {pages} {131601} (\bibinfo {year} {2004})},\ \Eprint {http://arxiv.org/abs/hep-ph/0310180} {arXiv:hep-ph/0310180} \BibitemShut {NoStop}%
\bibitem [{\citenamefont {Bevilacqua}\ \emph {et~al.}(2024)\citenamefont {Bevilacqua}, \citenamefont {Kowalski-Glikman},\ and\ \citenamefont {Wislicki}}]{Bevilacqua:2024jpy}%
  \BibitemOpen
  \bibfield  {author} {\bibinfo {author} {\bibfnamefont {A.}~\bibnamefont {Bevilacqua}}, \bibinfo {author} {\bibfnamefont {J.}~\bibnamefont {Kowalski-Glikman}}, \ and\ \bibinfo {author} {\bibfnamefont {W.}~\bibnamefont {Wislicki}},\ }\href@noop {} {\bibfield  {journal} {\bibinfo  {journal} {Phys. Rev. D in print.}\ } (\bibinfo {year} {2024})},\ \Eprint {http://arxiv.org/abs/2404.03600} {arXiv:2404.03600 [hep-ph]} \BibitemShut {NoStop}%
\bibitem [{\citenamefont {Gorini}\ \emph {et~al.}(1976)\citenamefont {Gorini}, \citenamefont {Kossakowski},\ and\ \citenamefont {Sudarshan}}]{Gorini:1975nb}%
  \BibitemOpen
  \bibfield  {author} {\bibinfo {author} {\bibfnamefont {V.}~\bibnamefont {Gorini}}, \bibinfo {author} {\bibfnamefont {A.}~\bibnamefont {Kossakowski}}, \ and\ \bibinfo {author} {\bibfnamefont {E.~C.~G.}\ \bibnamefont {Sudarshan}},\ }\href {\doibase 10.1063/1.522979} {\bibfield  {journal} {\bibinfo  {journal} {J. Math. Phys.}\ }\textbf {\bibinfo {volume} {17}},\ \bibinfo {pages} {821} (\bibinfo {year} {1976})}\BibitemShut {NoStop}%
\bibitem [{\citenamefont {Lindblad}(1976)}]{Lindblad:1975ef}%
  \BibitemOpen
  \bibfield  {author} {\bibinfo {author} {\bibfnamefont {G.}~\bibnamefont {Lindblad}},\ }\href {\doibase 10.1007/BF01608499} {\bibfield  {journal} {\bibinfo  {journal} {Commun. Math. Phys.}\ }\textbf {\bibinfo {volume} {48}},\ \bibinfo {pages} {119} (\bibinfo {year} {1976})}\BibitemShut {NoStop}%
\bibitem [{\citenamefont {Wald}(1980)}]{waldQuantumGravityTime1980}%
  \BibitemOpen
  \bibfield  {author} {\bibinfo {author} {\bibfnamefont {R.~M.}\ \bibnamefont {Wald}},\ }\href {\doibase 10.1103/PhysRevD.21.2742} {\bibfield  {journal} {\bibinfo  {journal} {Phys. Rev. D}\ }\textbf {\bibinfo {volume} {21}},\ \bibinfo {pages} {2742} (\bibinfo {year} {1980})}\BibitemShut {NoStop}%
\bibitem [{\citenamefont {Hawking}(1982)}]{hawkingUnpredictabilityQuantumGravity1982}%
  \BibitemOpen
  \bibfield  {author} {\bibinfo {author} {\bibfnamefont {S.~W.}\ \bibnamefont {Hawking}},\ }\href {\doibase 10.1007/BF01206031} {\bibfield  {journal} {\bibinfo  {journal} {Commun.Math. Phys.}\ }\textbf {\bibinfo {volume} {87}},\ \bibinfo {pages} {395} (\bibinfo {year} {1982})}\BibitemShut {NoStop}%
\bibitem [{\citenamefont {Ellis}\ \emph {et~al.}(1984)\citenamefont {Ellis}, \citenamefont {Hagelin}, \citenamefont {Nanopoulos},\ and\ \citenamefont {Srednicki}}]{Ellis:1983jz}%
  \BibitemOpen
  \bibfield  {author} {\bibinfo {author} {\bibfnamefont {J.~R.}\ \bibnamefont {Ellis}}, \bibinfo {author} {\bibfnamefont {J.~S.}\ \bibnamefont {Hagelin}}, \bibinfo {author} {\bibfnamefont {D.~V.}\ \bibnamefont {Nanopoulos}}, \ and\ \bibinfo {author} {\bibfnamefont {M.}~\bibnamefont {Srednicki}},\ }\href {\doibase 10.1016/0550-3213(84)90053-1} {\bibfield  {journal} {\bibinfo  {journal} {Nucl. Phys. B}\ }\textbf {\bibinfo {volume} {241}},\ \bibinfo {pages} {381} (\bibinfo {year} {1984})}\BibitemShut {NoStop}%
\bibitem [{\citenamefont {Huet}\ and\ \citenamefont {Peskin}(1995)}]{Huet:1994kr}%
  \BibitemOpen
  \bibfield  {author} {\bibinfo {author} {\bibfnamefont {P.}~\bibnamefont {Huet}}\ and\ \bibinfo {author} {\bibfnamefont {M.~E.}\ \bibnamefont {Peskin}},\ }\href {\doibase 10.1016/0550-3213(94)00390-Z} {\bibfield  {journal} {\bibinfo  {journal} {Nucl. Phys. B}\ }\textbf {\bibinfo {volume} {434}},\ \bibinfo {pages} {3} (\bibinfo {year} {1995})},\ \Eprint {http://arxiv.org/abs/hep-ph/9403257} {arXiv:hep-ph/9403257} \BibitemShut {NoStop}%
\bibitem [{\citenamefont {Kosteleck{\'y}}\ and\ \citenamefont {Russell}(2011)}]{kosteleckyDataTablesLorentz2011}%
  \BibitemOpen
  \bibfield  {author} {\bibinfo {author} {\bibfnamefont {V.~A.}\ \bibnamefont {Kosteleck{\'y}}}\ and\ \bibinfo {author} {\bibfnamefont {N.}~\bibnamefont {Russell}},\ }\href {\doibase 10.1103/RevModPhys.83.11} {\bibfield  {journal} {\bibinfo  {journal} {Rev. Mod. Phys.}\ }\textbf {\bibinfo {volume} {83}},\ \bibinfo {pages} {11} (\bibinfo {year} {2011})}\BibitemShut {NoStop}%
\bibitem [{\citenamefont {Link}\ \emph {et~al.}(2003)\citenamefont {Link} \emph {et~al.}}]{linkCharmSystemTests2003}%
  \BibitemOpen
  \bibfield  {author} {\bibinfo {author} {\bibfnamefont {J.}~\bibnamefont {Link}} \emph {et~al.} (\bibinfo {collaboration} {FOCUS Collaboration}),\ }\href {\doibase 10.1016/S0370-2693(03)00103-5} {\bibfield  {journal} {\bibinfo  {journal} {Physics Letters B}\ }\textbf {\bibinfo {volume} {556}},\ \bibinfo {pages} {7} (\bibinfo {year} {2003})}\BibitemShut {NoStop}%
\bibitem [{\citenamefont {Adam}\ \emph {et~al.}(2015)\citenamefont {Adam} \emph {et~al.}}]{alicecollaborationPrecisionMeasurementMass2015}%
  \BibitemOpen
  \bibfield  {author} {\bibinfo {author} {\bibfnamefont {J.}~\bibnamefont {Adam}} \emph {et~al.} (\bibinfo {collaboration} {ALICE Collaboration}),\ }\href {\doibase 10.1038/nphys3432} {\bibfield  {journal} {\bibinfo  {journal} {Nature Phys.}\ }\textbf {\bibinfo {volume} {11}},\ \bibinfo {pages} {811} (\bibinfo {year} {2015})},\ \Eprint {http://arxiv.org/abs/1508.03986} {arXiv:1508.03986 [nucl-ex]} \BibitemShut {NoStop}%
\bibitem [{\citenamefont {Bevan}(2016)}]{bevanTestingDiscreteSymmetries2016}%
  \BibitemOpen
  \bibfield  {author} {\bibinfo {author} {\bibfnamefont {A.~J.}\ \bibnamefont {Bevan}},\ }\href {\doibase 10.1007/s11467-015-0481-1} {\bibfield  {journal} {\bibinfo  {journal} {Front. Phys.}\ }\textbf {\bibinfo {volume} {11}},\ \bibinfo {pages} {111401} (\bibinfo {year} {2016})}\BibitemShut {NoStop}%
\bibitem [{\citenamefont {Moskal}\ \emph {et~al.}(2021)\citenamefont {Moskal} \emph {et~al.}}]{moskalTestingCPTSymmetry2021}%
  \BibitemOpen
  \bibfield  {author} {\bibinfo {author} {\bibfnamefont {P.}~\bibnamefont {Moskal}} \emph {et~al.} (\bibinfo {collaboration} {J-PET Collaboration}),\ }\href {\doibase 10.1038/s41467-021-25905-9} {\bibfield  {journal} {\bibinfo  {journal} {Nat Commun}\ }\textbf {\bibinfo {volume} {12}},\ \bibinfo {pages} {5658} (\bibinfo {year} {2021})}\BibitemShut {NoStop}%
\bibitem [{\citenamefont {Babusci}\ \emph {et~al.}(2023)\citenamefont {Babusci} \emph {et~al.}}]{babusciDirectTestsCP2023}%
  \BibitemOpen
  \bibfield  {author} {\bibinfo {author} {\bibfnamefont {D.}~\bibnamefont {Babusci}} \emph {et~al.} (\bibinfo {collaboration} {KLOE-2 Collaboration}),\ }\href {\doibase 10.1016/j.physletb.2023.138164} {\bibfield  {journal} {\bibinfo  {journal} {Physics Letters B}\ }\textbf {\bibinfo {volume} {845}},\ \bibinfo {pages} {138164} (\bibinfo {year} {2023})}\BibitemShut {NoStop}%
\bibitem [{\citenamefont {Caloni}\ \emph {et~al.}(2023)\citenamefont {Caloni}, \citenamefont {Giardiello}, \citenamefont {Lembo}, \citenamefont {Gerbino}, \citenamefont {Gubitosi}, \citenamefont {Lattanzi},\ and\ \citenamefont {Pagano}}]{caloniProbingLorentzviolatingElectrodynamics2023}%
  \BibitemOpen
  \bibfield  {author} {\bibinfo {author} {\bibfnamefont {L.}~\bibnamefont {Caloni}}, \bibinfo {author} {\bibfnamefont {S.}~\bibnamefont {Giardiello}}, \bibinfo {author} {\bibfnamefont {M.}~\bibnamefont {Lembo}}, \bibinfo {author} {\bibfnamefont {M.}~\bibnamefont {Gerbino}}, \bibinfo {author} {\bibfnamefont {G.}~\bibnamefont {Gubitosi}}, \bibinfo {author} {\bibfnamefont {M.}~\bibnamefont {Lattanzi}}, \ and\ \bibinfo {author} {\bibfnamefont {L.}~\bibnamefont {Pagano}},\ }\href {\doibase 10.1088/1475-7516/2023/03/018} {\bibfield  {journal} {\bibinfo  {journal} {J. Cosmol. Astropart. Phys.}\ }\textbf {\bibinfo {volume} {2023}},\ \bibinfo {pages} {018} (\bibinfo {year} {2023})}\BibitemShut {NoStop}%
\bibitem [{\citenamefont {Mishra}\ \emph {et~al.}(2024)\citenamefont {Mishra}, \citenamefont {Shukla}, \citenamefont {Singh},\ and\ \citenamefont {Singh}}]{mishraSearchLorentzviolationSidereal2023}%
  \BibitemOpen
  \bibfield  {author} {\bibinfo {author} {\bibfnamefont {S.}~\bibnamefont {Mishra}}, \bibinfo {author} {\bibfnamefont {S.}~\bibnamefont {Shukla}}, \bibinfo {author} {\bibfnamefont {L.}~\bibnamefont {Singh}}, \ and\ \bibinfo {author} {\bibfnamefont {V.}~\bibnamefont {Singh}},\ }\href {\doibase 10.1103/PhysRevD.109.075042} {\bibfield  {journal} {\bibinfo  {journal} {Phys. Rev. D}\ }\textbf {\bibinfo {volume} {109}},\ \bibinfo {pages} {075042} (\bibinfo {year} {2024})}\BibitemShut {NoStop}%
\bibitem [{\citenamefont {Karan}\ \emph {et~al.}(2018)\citenamefont {Karan}, \citenamefont {Nayak}, \citenamefont {Sinha},\ and\ \citenamefont {London}}]{karanUsingTimedependentIndirect2018}%
  \BibitemOpen
  \bibfield  {author} {\bibinfo {author} {\bibfnamefont {A.}~\bibnamefont {Karan}}, \bibinfo {author} {\bibfnamefont {A.~K.}\ \bibnamefont {Nayak}}, \bibinfo {author} {\bibfnamefont {R.}~\bibnamefont {Sinha}}, \ and\ \bibinfo {author} {\bibfnamefont {D.}~\bibnamefont {London}},\ }\href {\doibase 10.1016/j.physletb.2018.04.029} {\bibfield  {journal} {\bibinfo  {journal} {Physics Letters B}\ }\textbf {\bibinfo {volume} {781}},\ \bibinfo {pages} {459} (\bibinfo {year} {2018})}\BibitemShut {NoStop}%
\bibitem [{\citenamefont {Karan}(2020)}]{karanDealingCPTViolations2020}%
  \BibitemOpen
  \bibfield  {author} {\bibinfo {author} {\bibfnamefont {A.}~\bibnamefont {Karan}},\ }\href {\doibase 10.1140/epjc/s10052-020-8297-8} {\bibfield  {journal} {\bibinfo  {journal} {Eur. Phys. J. C}\ }\textbf {\bibinfo {volume} {80}},\ \bibinfo {pages} {782} (\bibinfo {year} {2020})}\BibitemShut {NoStop}%
\bibitem [{\citenamefont {Lehnert}(2022)}]{lehnertBetadecaySpectrumLorentz2022}%
  \BibitemOpen
  \bibfield  {author} {\bibinfo {author} {\bibfnamefont {R.}~\bibnamefont {Lehnert}},\ }\href {\doibase 10.1016/j.physletb.2022.137017} {\bibfield  {journal} {\bibinfo  {journal} {Physics Letters B}\ }\textbf {\bibinfo {volume} {828}},\ \bibinfo {pages} {137017} (\bibinfo {year} {2022})}\BibitemShut {NoStop}%
\bibitem [{\citenamefont {Vargas}(2024)}]{vargasProspectsTestingLorentz2024}%
  \BibitemOpen
  \bibfield  {author} {\bibinfo {author} {\bibfnamefont {A.~J.}\ \bibnamefont {Vargas}},\ }\href {\doibase 10.1103/PhysRevD.109.055001} {\bibfield  {journal} {\bibinfo  {journal} {Phys. Rev. D}\ }\textbf {\bibinfo {volume} {109}},\ \bibinfo {pages} {055001} (\bibinfo {year} {2024})}\BibitemShut {NoStop}%
\bibitem [{\citenamefont {Belyaev}\ \emph {et~al.}(2024)\citenamefont {Belyaev}, \citenamefont {Cerrito}, \citenamefont {Lunghi}, \citenamefont {Moretti},\ and\ \citenamefont {Sherrill}}]{belyaev2024probingcptinvariancequarks}%
  \BibitemOpen
  \bibfield  {author} {\bibinfo {author} {\bibfnamefont {A.}~\bibnamefont {Belyaev}}, \bibinfo {author} {\bibfnamefont {L.}~\bibnamefont {Cerrito}}, \bibinfo {author} {\bibfnamefont {E.}~\bibnamefont {Lunghi}}, \bibinfo {author} {\bibfnamefont {S.}~\bibnamefont {Moretti}}, \ and\ \bibinfo {author} {\bibfnamefont {N.}~\bibnamefont {Sherrill}},\ }\href {https://arxiv.org/abs/2405.12162} {\enquote {\bibinfo {title} {Probing cpt invariance with top quarks at the lhc},}\ } (\bibinfo {year} {2024}),\ \Eprint {http://arxiv.org/abs/2405.12162} {arXiv:2405.12162 [hep-ph]} \BibitemShut {NoStop}%
\bibitem [{\citenamefont {Nowak}\ \emph {et~al.}(2024)\citenamefont {Nowak}, \citenamefont {Malbrunot}, \citenamefont {Simon}, \citenamefont {Amsler}, \citenamefont {Cuendis}, \citenamefont {Lahs}, \citenamefont {Lanz}, \citenamefont {Nanda}, \citenamefont {Wiesinger}, \citenamefont {Wolz},\ and\ \citenamefont {Widmann}}]{nowak2024cptlorentzsymmetrytests}%
  \BibitemOpen
  \bibfield  {author} {\bibinfo {author} {\bibfnamefont {L.}~\bibnamefont {Nowak}}, \bibinfo {author} {\bibfnamefont {C.}~\bibnamefont {Malbrunot}}, \bibinfo {author} {\bibfnamefont {M.~C.}\ \bibnamefont {Simon}}, \bibinfo {author} {\bibfnamefont {C.}~\bibnamefont {Amsler}}, \bibinfo {author} {\bibfnamefont {S.~A.}\ \bibnamefont {Cuendis}}, \bibinfo {author} {\bibfnamefont {S.}~\bibnamefont {Lahs}}, \bibinfo {author} {\bibfnamefont {A.}~\bibnamefont {Lanz}}, \bibinfo {author} {\bibfnamefont {A.}~\bibnamefont {Nanda}}, \bibinfo {author} {\bibfnamefont {M.}~\bibnamefont {Wiesinger}}, \bibinfo {author} {\bibfnamefont {T.}~\bibnamefont {Wolz}}, \ and\ \bibinfo {author} {\bibfnamefont {E.}~\bibnamefont {Widmann}},\ }\href {https://arxiv.org/abs/2403.17763} {\enquote {\bibinfo {title} {Cpt and lorentz symmetry tests with hydrogen using a novel in-beam hyperfine spectroscopy method applicable to antihydrogen experiments},}\ } (\bibinfo {year} {2024}),\ \Eprint {http://arxiv.org/abs/2403.17763} {arXiv:2403.17763
  [hep-ex]} \BibitemShut {NoStop}%
\bibitem [{\citenamefont {Bernab{\'e}u}\ \emph {et~al.}(2006)\citenamefont {Bernab{\'e}u}, \citenamefont {Mavromatos},\ and\ \citenamefont {Sarkar}}]{bernabeuDecoherenceInducedViolation2006}%
  \BibitemOpen
  \bibfield  {author} {\bibinfo {author} {\bibfnamefont {J.}~\bibnamefont {Bernab{\'e}u}}, \bibinfo {author} {\bibfnamefont {N.~E.}\ \bibnamefont {Mavromatos}}, \ and\ \bibinfo {author} {\bibfnamefont {S.}~\bibnamefont {Sarkar}},\ }\href {\doibase 10.1103/PhysRevD.74.045014} {\bibfield  {journal} {\bibinfo  {journal} {Phys. Rev. D}\ }\textbf {\bibinfo {volume} {74}},\ \bibinfo {pages} {045014} (\bibinfo {year} {2006})}\BibitemShut {NoStop}%
\bibitem [{\citenamefont {Edwards}\ and\ \citenamefont {Kosteleck{\'y}}(2019)}]{edwardsSearchingCPTViolation2019}%
  \BibitemOpen
  \bibfield  {author} {\bibinfo {author} {\bibfnamefont {B.~R.}\ \bibnamefont {Edwards}}\ and\ \bibinfo {author} {\bibfnamefont {V.~A.}\ \bibnamefont {Kosteleck{\'y}}},\ }\href {\doibase 10.1016/j.physletb.2019.07.012} {\bibfield  {journal} {\bibinfo  {journal} {Physics Letters B}\ }\textbf {\bibinfo {volume} {795}},\ \bibinfo {pages} {620} (\bibinfo {year} {2019})}\BibitemShut {NoStop}%
\bibitem [{\citenamefont {Huang}\ and\ \citenamefont {Shi}(2014)}]{huangViolatingParametersDetermined2014}%
  \BibitemOpen
  \bibfield  {author} {\bibinfo {author} {\bibfnamefont {Z.}~\bibnamefont {Huang}}\ and\ \bibinfo {author} {\bibfnamefont {Y.}~\bibnamefont {Shi}},\ }\href {\doibase 10.1103/PhysRevD.89.016018} {\bibfield  {journal} {\bibinfo  {journal} {Phys. Rev. D}\ }\textbf {\bibinfo {volume} {89}},\ \bibinfo {pages} {016018} (\bibinfo {year} {2014})}\BibitemShut {NoStop}%
\bibitem [{\citenamefont {Shi}(2013)}]{shiExactResultsCP2013}%
  \BibitemOpen
  \bibfield  {author} {\bibinfo {author} {\bibfnamefont {Y.}~\bibnamefont {Shi}},\ }\href {\doibase 10.1140/epjc/s10052-013-2506-7} {\bibfield  {journal} {\bibinfo  {journal} {Eur. Phys. J. C}\ }\textbf {\bibinfo {volume} {73}},\ \bibinfo {pages} {2506} (\bibinfo {year} {2013})}\BibitemShut {NoStop}%
\bibitem [{\citenamefont {Shi}(2012)}]{shiExactTheoremsConcerning2012}%
  \BibitemOpen
  \bibfield  {author} {\bibinfo {author} {\bibfnamefont {Y.}~\bibnamefont {Shi}},\ }\href {\doibase 10.1140/epjc/s10052-012-1907-3} {\bibfield  {journal} {\bibinfo  {journal} {Eur. Phys. J. C}\ }\textbf {\bibinfo {volume} {72}},\ \bibinfo {pages} {1907} (\bibinfo {year} {2012})}\BibitemShut {NoStop}%
\bibitem [{\citenamefont {Kosteleck{\'y}}(1998)}]{kosteleckySensitivityCPTTests1998}%
  \BibitemOpen
  \bibfield  {author} {\bibinfo {author} {\bibfnamefont {V.~A.}\ \bibnamefont {Kosteleck{\'y}}},\ }\href {\doibase 10.1103/PhysRevLett.80.1818} {\bibfield  {journal} {\bibinfo  {journal} {Phys. Rev. Lett.}\ }\textbf {\bibinfo {volume} {80}},\ \bibinfo {pages} {1818} (\bibinfo {year} {1998})}\BibitemShut {NoStop}%
\bibitem [{\citenamefont {Kosteleck{\'y}}(1999)}]{kosteleckySignalsCPTLorentz1999}%
  \BibitemOpen
  \bibfield  {author} {\bibinfo {author} {\bibfnamefont {V.~A.}\ \bibnamefont {Kosteleck{\'y}}},\ }\href {\doibase 10.1103/PhysRevD.61.016002} {\bibfield  {journal} {\bibinfo  {journal} {Phys. Rev. D}\ }\textbf {\bibinfo {volume} {61}},\ \bibinfo {pages} {016002} (\bibinfo {year} {1999})}\BibitemShut {NoStop}%
\bibitem [{\citenamefont {Roberts}(2017)}]{robertsTestingSymmetryCorrelated2017}%
  \BibitemOpen
  \bibfield  {author} {\bibinfo {author} {\bibfnamefont {{\'A}.}~\bibnamefont {Roberts}},\ }\href {\doibase 10.1103/PhysRevD.96.116015} {\bibfield  {journal} {\bibinfo  {journal} {Phys. Rev. D}\ }\textbf {\bibinfo {volume} {96}},\ \bibinfo {pages} {116015} (\bibinfo {year} {2017})}\BibitemShut {NoStop}%
\bibitem [{\citenamefont {Aaij}\ \emph {et~al.}(2021)\citenamefont {Aaij} \emph {et~al.}}]{aaijSearchTimedependentViolation2021}%
  \BibitemOpen
  \bibfield  {author} {\bibinfo {author} {\bibfnamefont {R.}~\bibnamefont {Aaij}} \emph {et~al.} (\bibinfo {collaboration} {LHCb Collaboration}),\ }\href {\doibase 10.1103/PhysRevD.104.072010} {\bibfield  {journal} {\bibinfo  {journal} {Phys. Rev. D}\ }\textbf {\bibinfo {volume} {104}},\ \bibinfo {pages} {072010} (\bibinfo {year} {2021})}\BibitemShut {NoStop}%
\bibitem [{\citenamefont {et~al. {Particle Data Group}}(2022)}]{particledatagroupReviewParticlePhysics2022}%
  \BibitemOpen
  \bibfield  {author} {\bibinfo {author} {\bibfnamefont {R.~L.~W.}\ \bibnamefont {et~al. {Particle Data Group}}},\ }\href {\doibase 10.1093/ptep/ptac097} {\bibfield  {journal} {\bibinfo  {journal} {Progress of Theoretical and Experimental Physics}\ }\textbf {\bibinfo {volume} {2022}},\ \bibinfo {pages} {083C01} (\bibinfo {year} {2022})}\BibitemShut {NoStop}%
\bibitem [{\citenamefont {Weisskopf}\ and\ \citenamefont {Wigner}(1930)}]{weisskopfBerechnungNaturlichenLinienbreite1930}%
  \BibitemOpen
  \bibfield  {author} {\bibinfo {author} {\bibfnamefont {V.}~\bibnamefont {Weisskopf}}\ and\ \bibinfo {author} {\bibfnamefont {E.}~\bibnamefont {Wigner}},\ }\href {\doibase 10.1007/BF01336768} {\bibfield  {journal} {\bibinfo  {journal} {Z. Physik}\ }\textbf {\bibinfo {volume} {63}},\ \bibinfo {pages} {54} (\bibinfo {year} {1930})}\BibitemShut {NoStop}%
\bibitem [{\citenamefont {Sozzi}(2008)}]{sozzimarcos.DiscreteSymmetriesCP2008}%
  \BibitemOpen
  \bibfield  {author} {\bibinfo {author} {\bibfnamefont {M.~S.}\ \bibnamefont {Sozzi}},\ }\href@noop {} {\emph {\bibinfo {title} {{Discrete Symmetries and CP Violation: From Experiment to Theory}}}}\ (\bibinfo  {publisher} {Oxford University Press},\ \bibinfo {year} {2008})\BibitemShut {NoStop}%
\bibitem [{\citenamefont {Alan~Kosteleck{\'y}}\ and\ \citenamefont {Potting}(1996)}]{alankosteleckyExpectationValuesLorentz1996}%
  \BibitemOpen
  \bibfield  {author} {\bibinfo {author} {\bibfnamefont {V.}~\bibnamefont {Alan~Kosteleck{\'y}}}\ and\ \bibinfo {author} {\bibfnamefont {R.}~\bibnamefont {Potting}},\ }\href {\doibase 10.1016/0370-2693(96)00589-8} {\bibfield  {journal} {\bibinfo  {journal} {Physics Letters B}\ }\textbf {\bibinfo {volume} {381}},\ \bibinfo {pages} {89} (\bibinfo {year} {1996})}\BibitemShut {NoStop}%
\bibitem [{\citenamefont {Kosteleck{\'y}}(2001)}]{kosteleckyCPTLorentzViolation2001}%
  \BibitemOpen
  \bibfield  {author} {\bibinfo {author} {\bibfnamefont {V.~A.}\ \bibnamefont {Kosteleck{\'y}}},\ }\href {\doibase 10.1103/PhysRevD.64.076001} {\bibfield  {journal} {\bibinfo  {journal} {Phys. Rev. D}\ }\textbf {\bibinfo {volume} {64}},\ \bibinfo {pages} {076001} (\bibinfo {year} {2001})}\BibitemShut {NoStop}%
\bibitem [{\citenamefont {Van~Tilburg}\ and\ \citenamefont {Van~Veghel}(2015)}]{vantilburgStatusProspectsCPT2015}%
  \BibitemOpen
  \bibfield  {author} {\bibinfo {author} {\bibfnamefont {J.}~\bibnamefont {Van~Tilburg}}\ and\ \bibinfo {author} {\bibfnamefont {M.}~\bibnamefont {Van~Veghel}},\ }\href {\doibase 10.1016/j.physletb.2015.01.036} {\bibfield  {journal} {\bibinfo  {journal} {Physics Letters B}\ }\textbf {\bibinfo {volume} {742}},\ \bibinfo {pages} {236} (\bibinfo {year} {2015})}\BibitemShut {NoStop}%
\bibitem [{\citenamefont {Amhis}\ \emph {et~al.}(2023)\citenamefont {Amhis} \emph {et~al.}}]{amhisAveragesHadronHadron2023}%
  \BibitemOpen
  \bibfield  {author} {\bibinfo {author} {\bibfnamefont {Y.}~\bibnamefont {Amhis}} \emph {et~al.} (\bibinfo {collaboration} {HFLAV}),\ }\href {\doibase 10.1103/PhysRevD.107.052008} {\bibfield  {journal} {\bibinfo  {journal} {Phys. Rev. D}\ }\textbf {\bibinfo {volume} {107}},\ \bibinfo {pages} {052008} (\bibinfo {year} {2023})}\BibitemShut {NoStop}%
\bibitem [{\citenamefont {Pajero}\ and\ \citenamefont {Morello}(2022)}]{pajeroMixingCPViolation2022}%
  \BibitemOpen
  \bibfield  {author} {\bibinfo {author} {\bibfnamefont {T.}~\bibnamefont {Pajero}}\ and\ \bibinfo {author} {\bibfnamefont {M.~J.}\ \bibnamefont {Morello}},\ }\href {\doibase 10.1007/JHEP03(2022)162} {\bibfield  {journal} {\bibinfo  {journal} {J. High Energ. Phys.}\ }\textbf {\bibinfo {volume} {2022}},\ \bibinfo {pages} {162} (\bibinfo {year} {2022})}\BibitemShut {NoStop}%
\bibitem [{\citenamefont {Schwartz}(2022)}]{schwartzEffectBarMixing2022}%
  \BibitemOpen
  \bibfield  {author} {\bibinfo {author} {\bibfnamefont {A.~J.}\ \bibnamefont {Schwartz}},\ }\href@noop {} {} (\bibinfo {year} {2022}),\ \Eprint {http://arxiv.org/abs/2207.11867} {arxiv:2207.11867 [hep-ex, physics:hep-ph]} \BibitemShut {NoStop}%
\bibitem [{\citenamefont {Aaij}\ \emph {et~al.}(2019)\citenamefont {Aaij} \emph {et~al.}}]{aaijObservationMassDifference2019}%
  \BibitemOpen
  \bibfield  {author} {\bibinfo {author} {\bibfnamefont {R.}~\bibnamefont {Aaij}} \emph {et~al.} (\bibinfo {collaboration} {LHCb Collaboration}),\ }\href {\doibase 10.1103/PhysRevLett.122.211803} {\bibfield  {journal} {\bibinfo  {journal} {Phys. Rev. Lett.}\ }\textbf {\bibinfo {volume} {122}},\ \bibinfo {pages} {211803} (\bibinfo {year} {2019})}\BibitemShut {NoStop}%
\bibitem [{\citenamefont {Aaij}\ \emph {et~al.}(2020)\citenamefont {Aaij} \emph {et~al.}}]{aaijUpdatedmeasurment2020}%
  \BibitemOpen
  \bibfield  {author} {\bibinfo {author} {\bibfnamefont {R.}~\bibnamefont {Aaij}} \emph {et~al.} (\bibinfo {collaboration} {LHCb Collaboration}),\ }\href {\doibase 10.1103/PhysRevD.101.012005} {\bibfield  {journal} {\bibinfo  {journal} {Phys. Rev. D}\ }\textbf {\bibinfo {volume} {101}},\ \bibinfo {pages} {012005} (\bibinfo {year} {2020})}\BibitemShut {NoStop}%
\bibitem [{\citenamefont {Aaij}\ \emph {et~al.}(2023{\natexlab{a}})\citenamefont {Aaij} \emph {et~al.}}]{aaijTimeIntegrated2023}%
  \BibitemOpen
  \bibfield  {author} {\bibinfo {author} {\bibfnamefont {R.}~\bibnamefont {Aaij}} \emph {et~al.} (\bibinfo {collaboration} {LHCb Collaboration}),\ }\href {\doibase 10.1103/PhysRevLett.131.091802} {\bibfield  {journal} {\bibinfo  {journal} {Phys. Rev. Lett.}\ }\textbf {\bibinfo {volume} {131}},\ \bibinfo {pages} {091802} (\bibinfo {year} {2023}{\natexlab{a}})}\BibitemShut {NoStop}%
\bibitem [{\citenamefont {Angelopoulos}\ \emph {et~al.}(1999)\citenamefont {Angelopoulos} \emph {et~al.}}]{angelopoulos1999}%
  \BibitemOpen
  \bibfield  {author} {\bibinfo {author} {\bibfnamefont {A.}~\bibnamefont {Angelopoulos}} \emph {et~al.} (\bibinfo {collaboration} {CPLEAR Collaboration}),\ }\href {\doibase 10.1016/S0370-2693(99)01333-7} {\bibfield  {journal} {\bibinfo  {journal} {Phys. Lett. B}\ }\textbf {\bibinfo {volume} {471}},\ \bibinfo {pages} {332} (\bibinfo {year} {1999})}\BibitemShut {NoStop}%
\bibitem [{\citenamefont {Higuchi}\ \emph {et~al.}(2012)\citenamefont {Higuchi} \emph {et~al.}}]{Higuchi:2012kx}%
  \BibitemOpen
  \bibfield  {author} {\bibinfo {author} {\bibfnamefont {T.}~\bibnamefont {Higuchi}} \emph {et~al.},\ }\href {\doibase 10.1103/PhysRevD.85.071105} {\bibfield  {journal} {\bibinfo  {journal} {Phys. Rev. D}\ }\textbf {\bibinfo {volume} {85}},\ \bibinfo {pages} {071105} (\bibinfo {year} {2012})},\ \Eprint {http://arxiv.org/abs/1203.0930} {arXiv:1203.0930 [hep-ex]} \BibitemShut {NoStop}%
\bibitem [{\citenamefont {Lees}\ \emph {et~al.}(2016)\citenamefont {Lees} \emph {et~al.}}]{BaBar:2016zvy}%
  \BibitemOpen
  \bibfield  {author} {\bibinfo {author} {\bibfnamefont {J.~P.}\ \bibnamefont {Lees}} \emph {et~al.} (\bibinfo {collaboration} {BaBar Collaboration}),\ }\href {\doibase 10.1103/PhysRevD.94.011101} {\bibfield  {journal} {\bibinfo  {journal} {Phys. Rev. D}\ }\textbf {\bibinfo {volume} {94}},\ \bibinfo {pages} {011101} (\bibinfo {year} {2016})},\ \Eprint {http://arxiv.org/abs/1605.04545} {arXiv:1605.04545 [hep-ex]} \BibitemShut {NoStop}%
\bibitem [{\citenamefont {Aaij}\ \emph {et~al.}(2016)\citenamefont {Aaij} \emph {et~al.}}]{aaijSearchViolationsLorentz2016}%
  \BibitemOpen
  \bibfield  {author} {\bibinfo {author} {\bibfnamefont {R.}~\bibnamefont {Aaij}} \emph {et~al.} (\bibinfo {collaboration} {LHCb Collaboration}),\ }\href {\doibase 10.1103/PhysRevLett.116.241601} {\bibfield  {journal} {\bibinfo  {journal} {Phys. Rev. Lett.}\ }\textbf {\bibinfo {volume} {116}},\ \bibinfo {pages} {241601} (\bibinfo {year} {2016})}\BibitemShut {NoStop}%
\bibitem [{\citenamefont {Aaij}\ \emph {et~al.}(2018)\citenamefont {Aaij} \emph {et~al.}}]{LHCb:2018roe}%
  \BibitemOpen
  \bibfield  {author} {\bibinfo {author} {\bibfnamefont {R.}~\bibnamefont {Aaij}} \emph {et~al.} (\bibinfo {collaboration} {LHCb Collaboration}),\ }\href@noop {} {\emph {\bibinfo {title} {{Physics case for an LHCb Upgrade II - Opportunities in flavour physics, and beyond, in the HL-LHC era}}}},\ \bibinfo {type} {Tech. Rep.}\ (\bibinfo  {institution} {LHCb},\ \bibinfo {year} {2018})\ \Eprint {http://arxiv.org/abs/1808.08865} {arXiv:1808.08865 [hep-ex]} \BibitemShut {NoStop}%
\bibitem [{\citenamefont {Aaij}\ \emph {et~al.}(2023{\natexlab{b}})\citenamefont {Aaij} \emph {et~al.}}]{run3fig}%
  \BibitemOpen
  \bibfield  {author} {\bibinfo {author} {\bibfnamefont {R.}~\bibnamefont {Aaij}} \emph {et~al.} (\bibinfo {collaboration} {LHCb Collaboration}),\ }\href@noop {} {}\bibinfo {howpublished} {LHCb-FIGURE-2023-011} (\bibinfo {year} {2023}{\natexlab{b}})\BibitemShut {NoStop}%
\bibitem [{\citenamefont {Shi}\ and\ \citenamefont {Yang}(2018)}]{Shi:2016bvo}%
  \BibitemOpen
  \bibfield  {author} {\bibinfo {author} {\bibfnamefont {Y.}~\bibnamefont {Shi}}\ and\ \bibinfo {author} {\bibfnamefont {J.}~\bibnamefont {Yang}},\ }\href {\doibase 10.1103/PhysRevD.98.075019} {\bibfield  {journal} {\bibinfo  {journal} {Phys. Rev. D}\ }\textbf {\bibinfo {volume} {98}},\ \bibinfo {pages} {075019} (\bibinfo {year} {2018})},\ \Eprint {http://arxiv.org/abs/1612.07628} {arXiv:1612.07628 [hep-ph]} \BibitemShut {NoStop}%
\bibitem [{\citenamefont {Banuls}\ and\ \citenamefont {Bernabeu}(1999)}]{Banuls:1999aj}%
  \BibitemOpen
  \bibfield  {author} {\bibinfo {author} {\bibfnamefont {M.~C.}\ \bibnamefont {Banuls}}\ and\ \bibinfo {author} {\bibfnamefont {J.}~\bibnamefont {Bernabeu}},\ }\href {\doibase 10.1016/S0370-2693(99)01043-6} {\bibfield  {journal} {\bibinfo  {journal} {Phys. Lett. B}\ }\textbf {\bibinfo {volume} {464}},\ \bibinfo {pages} {117} (\bibinfo {year} {1999})},\ \Eprint {http://arxiv.org/abs/hep-ph/9908353} {arXiv:hep-ph/9908353} \BibitemShut {NoStop}%
\bibitem [{\citenamefont {Bernabeu}\ \emph {et~al.}(2012)\citenamefont {Bernabeu}, \citenamefont {Martinez-Vidal},\ and\ \citenamefont {Villanueva-Perez}}]{Bernabeu:2012ab}%
  \BibitemOpen
  \bibfield  {author} {\bibinfo {author} {\bibfnamefont {J.}~\bibnamefont {Bernabeu}}, \bibinfo {author} {\bibfnamefont {F.}~\bibnamefont {Martinez-Vidal}}, \ and\ \bibinfo {author} {\bibfnamefont {P.}~\bibnamefont {Villanueva-Perez}},\ }\href {\doibase 10.1007/JHEP08(2012)064} {\bibfield  {journal} {\bibinfo  {journal} {JHEP}\ }\textbf {\bibinfo {volume} {08}},\ \bibinfo {pages} {064} (\bibinfo {year} {2012})},\ \Eprint {http://arxiv.org/abs/1203.0171} {arXiv:1203.0171 [hep-ph]} \BibitemShut {NoStop}%
\bibitem [{\citenamefont {Bernabeu}\ \emph {et~al.}(2013)\citenamefont {Bernabeu}, \citenamefont {Di~Domenico},\ and\ \citenamefont {Villanueva-Perez}}]{Bernabeu:2012nu}%
  \BibitemOpen
  \bibfield  {author} {\bibinfo {author} {\bibfnamefont {J.}~\bibnamefont {Bernabeu}}, \bibinfo {author} {\bibfnamefont {A.}~\bibnamefont {Di~Domenico}}, \ and\ \bibinfo {author} {\bibfnamefont {P.}~\bibnamefont {Villanueva-Perez}},\ }\href {\doibase 10.1016/j.nuclphysb.2012.11.009} {\bibfield  {journal} {\bibinfo  {journal} {Nucl. Phys. B}\ }\textbf {\bibinfo {volume} {868}},\ \bibinfo {pages} {102} (\bibinfo {year} {2013})},\ \Eprint {http://arxiv.org/abs/1208.0773} {arXiv:1208.0773 [hep-ph]} \BibitemShut {NoStop}%
\bibitem [{\citenamefont {Bernabeu}\ \emph {et~al.}(2015)\citenamefont {Bernabeu}, \citenamefont {Di~Domenico},\ and\ \citenamefont {Villanueva-Perez}}]{Bernabeu:2015aga}%
  \BibitemOpen
  \bibfield  {author} {\bibinfo {author} {\bibfnamefont {J.}~\bibnamefont {Bernabeu}}, \bibinfo {author} {\bibfnamefont {A.}~\bibnamefont {Di~Domenico}}, \ and\ \bibinfo {author} {\bibfnamefont {P.}~\bibnamefont {Villanueva-Perez}},\ }\href {\doibase 10.1007/JHEP10(2015)139} {\bibfield  {journal} {\bibinfo  {journal} {JHEP}\ }\textbf {\bibinfo {volume} {10}},\ \bibinfo {pages} {139} (\bibinfo {year} {2015})},\ \Eprint {http://arxiv.org/abs/1509.02000} {arXiv:1509.02000 [hep-ph]} \BibitemShut {NoStop}%
\bibitem [{\citenamefont {Lees}\ \emph {et~al.}(2012)\citenamefont {Lees} \emph {et~al.}}]{BaBar:2012bwc}%
  \BibitemOpen
  \bibfield  {author} {\bibinfo {author} {\bibfnamefont {J.~P.}\ \bibnamefont {Lees}} \emph {et~al.} (\bibinfo {collaboration} {BaBar Collaboration}),\ }\href {\doibase 10.1103/PhysRevLett.109.211801} {\bibfield  {journal} {\bibinfo  {journal} {Phys. Rev. Lett.}\ }\textbf {\bibinfo {volume} {109}},\ \bibinfo {pages} {211801} (\bibinfo {year} {2012})},\ \Eprint {http://arxiv.org/abs/1207.5832} {arXiv:1207.5832 [hep-ex]} \BibitemShut {NoStop}%
\bibitem [{\citenamefont {Ablikim}\ \emph {et~al.}(2010)\citenamefont {Ablikim} \emph {et~al.}}]{BESIII:2009fln}%
  \BibitemOpen
  \bibfield  {author} {\bibinfo {author} {\bibfnamefont {M.}~\bibnamefont {Ablikim}} \emph {et~al.} (\bibinfo {collaboration} {BESIII}),\ }\href {\doibase 10.1016/j.nima.2009.12.050} {\bibfield  {journal} {\bibinfo  {journal} {Nucl. Instrum. Meth. A}\ }\textbf {\bibinfo {volume} {614}},\ \bibinfo {pages} {345} (\bibinfo {year} {2010})},\ \Eprint {http://arxiv.org/abs/0911.4960} {arXiv:0911.4960 [physics.ins-det]} \BibitemShut {NoStop}%
\bibitem [{\citenamefont {Achasov}\ \emph {et~al.}(2024)\citenamefont {Achasov} \emph {et~al.}}]{Achasov:2023gey}%
  \BibitemOpen
  \bibfield  {author} {\bibinfo {author} {\bibfnamefont {M.}~\bibnamefont {Achasov}} \emph {et~al.},\ }\href {\doibase 10.1007/s11467-023-1333-z} {\bibfield  {journal} {\bibinfo  {journal} {Front. Phys. (Beijing)}\ }\textbf {\bibinfo {volume} {19}},\ \bibinfo {pages} {14701} (\bibinfo {year} {2024})},\ \Eprint {http://arxiv.org/abs/2303.15790} {arXiv:2303.15790 [hep-ex]} \BibitemShut {NoStop}%
\bibitem [{\citenamefont {Naik}(2023)}]{naikNovelCorrelatedOverline2023}%
  \BibitemOpen
  \bibfield  {author} {\bibinfo {author} {\bibfnamefont {P.}~\bibnamefont {Naik}},\ }\href {\doibase 10.1007/JHEP03(2023)038} {\bibfield  {journal} {\bibinfo  {journal} {J. High Energ. Phys.}\ }\textbf {\bibinfo {volume} {2023}},\ \bibinfo {pages} {38} (\bibinfo {year} {2023})}\BibitemShut {NoStop}%
\end{thebibliography}%

\appendix
\section{Relation between the CPT linear terms and $\delta m$, $\delta \Gamma$ parameters}
\label{appendix:relation}
Let's consider the linear terms of the RS asymmetry corresponding to the slope
in Eq. (\ref{eq:ACPTlinear}) $s = \text{Re}{(z)}y-\text{Im}{(z)}x$. 
We derive the expression on the $\delta m$ and $\delta \Gamma$ parameters. The CPT-violating parameter $z$ can be expressed as
\begin{equation}
\label{eq:zvsmassdiff}
z = \frac{\delta m - i \delta \Gamma/ 2 }{\Delta \lambda}=\frac{\delta m - i \delta \Gamma/ 2 }{\Delta m - \Delta \Gamma /2},
\end{equation}
Combining Eqs. (\ref{eq:zvsmassdiff}) and (\ref{eq:xy_def}), we obtain the following relation between $z$ and $\delta m$ as well as $\delta \Gamma$.

\begin{equation}
z=\frac{\delta m - i \delta \Gamma/2}{\Gamma(x-iy)}=\frac{(x+iy)(\delta m - i \delta \Gamma/2)}{\Gamma(x^2+y^2)}.
\label{a1}
\end{equation}

\noindent Taking the real and imaginary values of expression (\ref{a1}), we obtain:

\begin{equation}
\text{Re}{(z)}=\frac{x\delta m + y \delta \Gamma/2}{\Gamma(x^2+y^2)}.
\label{a2}
\end{equation}

\begin{equation}
\text{Im}{(z)}=\frac{y \delta m - x \delta \Gamma/2}{\Gamma(x^2+y^2)}.
\label{a3}
\end{equation}

\noindent. Let us write the expression $\text{Re}{(z)}y-\text{Im}{(z)}x$ using Eqs. (\ref{a2}) and (\ref{a3}):

\begin{equation}
\begin{aligned}
&s=\text{Re}(z)y - \text{Im}(z)x \\
&= \frac{(yx\delta m + y^2\delta \Gamma/2)}{\Gamma(x^2+y^2)} - \frac{(yx\delta m - x^2\delta \Gamma/2)}{\Gamma(x^2+y^2)} \\
&= \frac{\delta \Gamma}{2\Gamma}.
\end{aligned}
\label{a4}
\end{equation}
The slope of the asymmetry is directly related to the decay width difference $\frac{\delta \Gamma}{2\Gamma} $.

\noindent To express the relation for $\delta m$ an additional assumption, for instance, about the $z$ phase $\theta \equiv arg(z)$  is needed . 
Let's start with the case $\text{Re}(z)=0 \leftrightarrow \theta = \frac{\pi}{2},\frac{3\pi}{2} $.
This leads to the following expression:
\begin{equation}
\text{Re}{(z)}=0 \Rightarrow  \frac{\delta m}{\Gamma} = -\frac{y}{x} \frac{\delta \Gamma}{2\Gamma} = -\frac{y}{x}  s
\end{equation}
If $\text{Re}(z) \ne 0$, then using Eqs. (\ref{a2}), (\ref{a3}) we can express  
$\tan(\theta)$ as:

\begin{equation}
\tan(\theta) = \frac{\text{Im}{(z)}}{\text{Re}{(z)}} = \frac{y \delta m - x \delta\Gamma/2}{x \delta m + y \delta\Gamma/2}.
\label{tgphi}
\end{equation}
Using the Eq. (\ref{tgphi})  $\delta m$ can be expressed:
\begin{equation}
\frac{\delta m}{\Gamma}(\theta) = \frac{\delta \Gamma}{2 \Gamma} \times \frac{x + y  \tan(\theta)}{y -x \tan(\theta)} = s \times \frac{x + y  \tan(\theta)}{y -x \tan(\theta)} 
\label{deltamphi}
\end{equation}
Eq. (\ref{deltamphi}) has an undefined value for $y = \tan(\theta) x$, which corresponds to the situation, where the linear term vanishes, and the higher order terms of the RS asymmetry must be taken into account. This is a case of the SME framework~\cite{kosteleckyCPTLorentzViolation2001}.

Finally,  for the particular case of $\theta = 0, \pi \leftrightarrow \text{Im}(z)=0$ we recover the solution:
\begin{equation}
\frac{\delta m}{\Gamma} =  s \times \frac{x}{y} 
\end{equation}

\end{document}